\documentclass[11pt]{article}
\usepackage{graphics,epsfig}
\usepackage[psamsfonts]{amssymb}

\newtheorem{definition}{Definition}

\newtheorem{proposition}{Proposition}

\def\orthrel{\hbox{$\perp$ \hskip - 11.1 pt \raise 3.2pt
\hbox{$\scriptstyle \wedge$}}\ }
\newcommand{\be}{\begin{equation}}
\newcommand{\ee}{\end{equation}}
\newcommand{\bea}{\begin{eqnarray}}
\newcommand{\eea}{\end{eqnarray}}
\newcommand{\ba}{\begin{array}}
\newcommand{\ea}{\end{array}}
\newcommand{\bd}{\begin{definition}}
\newcommand{\ed}{\end{definition}}
\newcommand{\bp}{\begin{proposition}}
\newcommand{\ep}{\end{proposition}}
\newcommand{\compl}{{\mathbb C}}

\begin{document}
\title{Contextualizing Concepts using a Mathematical \\ Generalization of
the Quantum Formalism\footnote{Published as: Gabora, L. and Aerts, D. (2002). Contextualizing concepts using a mathematical generalization of
the quantum formalism. {\it Journal of Experimental and Theoretical Artificial Intelligence}, {\bf 14}, pp. 327-358.}}
\author{Liane Gabora and Diederik Aerts\\
        \normalsize\itshape
        Center Leo Apostel for Interdisciplinary Studies (CLEA) \\
        \normalsize\itshape
        Vrije Universiteit Brussel, Krijgskundestraat 33 \\
		\normalsize\itshape
		1160 Brussels, Belgium \\
        \normalsize
        E-Mails: \textsf{lgabora@vub.ac.be, diraerts@vub.ac.be} \\
}
\date{}
\maketitle
\begin{abstract}
\noindent
We outline the rationale and preliminary results of using the State Context
Property (SCOP) formalism, originally developed as
a generalization of quantum mechanics, to describe the contextual manner in
which concepts are evoked, used, and combined to
generate meaning. The quantum formalism was developed to cope with problems
arising in the description of (1) the measurement
process, and (2) the generation of new states with new properties when
particles become entangled. Similar problems arising
with concepts motivated the formal treatment introduced here. Concepts are
viewed not as fixed representations, but entities
existing in states of potentiality that require interaction with a
context---a stimulus or another concept---to `collapse' to an instantiated
form ({\it e.g.} exemplar, prototype, or other possibly imaginary
instance). The stimulus situation plays the role of
the measurement in physics, acting as context that induces a change of the
cognitive state from
superposition state to collapsed state. The collapsed state is
more likely to consist of a conjunction of
concepts for associative than analytic thought because more stimulus or
concept properties take part in the
collapse. We provide two contextual measures of conceptual distance---one
using collapse probabilities and the other weighted
properties---and show how they can be applied to conjunctions using the pet
fish problem.
\end{abstract}

\section{Introduction}
Theories of concepts have by and large been {\it representational
theories}. By this we mean that
concepts are seen to take the form of fixed mental representations, as
opposed to being
constructed, or `re-constructed', on the fly through the interaction
between the cognitive state and the situation or context.

Representational theories have met with some success. They are adequate for
predicting
experimental results for many dependent variables including typicality
ratings, latency of
category decision, exemplar generation frequencies, and category naming
frequencies. However
increasingly, for both theoretical and empirical reasons, they are coming
under fire ({\it e.g.}
Riegler, Peschl and von Stein 1999, Rosch 1999). As Rosch puts it, they do
not account for the
fact that concepts have a participatory, not an identifying function in
situations. That is,
they cannot explain the contextual manner in which concepts are evoked and
used (see also Murphy
and Medin 1985, Hampton 1987, Medin and Shoben 1988, Gerrig
and Murphy 1992, Komatsu 1992).
Contextuality is the reason why representational theories cannot describe
or predict what
happens when two or more concepts arise together, or follow one another, as
in the creative
generation or interpretation of {\it conjunctions} of concepts.

This paper shows how formalisms designed to cope with context and
conjunction in the
microworld may be a source of inspiration for a description of concepts. In
this {\it contextualized theory}\footnote{Not to be confused with the {\it
context model} (Medin and Schaffer 1978, Nosofsky 1986), an exemplar-based
representational theory. Whereas in the context model, a concept is viewed
as a static representation derived from context-specific
concretely-experienced instances, in our
approach, a concept is viewed as a predisposition to dynamically attract
context-specific cognitive states (both concrete stimulus experiences and
imagined or counterfactual situations) into a certain subspace of
conceptual space.},
not only does a concept give meaning to a stimulus or situation, but the
situation evokes meaning in the concept,
and when more than one is active they evoke meaning in each other.

\section{Limitations of Representational Approaches}
\label{sec:traditionaltheories}
We begin by briefly summarizing some of the most influential representational
theories of concepts, and efforts to delineate what a concept is with the
notion of conceptual distance. We then discuss difficulties encountered
with representational approaches in predicting membership
assessment for conjunctions of concepts. We then show that representational
theories have even more trouble coping with the spontaneous emergence or
loss of features that can occur when concepts combine.

\subsection{Theories of Concepts and Conceptual Distance}
According to the {\it classical theory} of concepts, there exists for each
concept a set of
defining features that are singly necessary and jointly sufficient ({\it
e.g}. Sutcliffe 1993).
Extensive evidence has been provided against this theory (for overviews see
Smith
and Medin 1981, Komatsu 1992).

A number of alternatives have been put forth. According to the {\it prototype
theory} (Rosch 1975a,
1978, 1983, Rosch and Mervis 1975), concepts are represented by a set of,
not {\it defining}, but
{\it characteristic} features, which are weighted in the definition of the
prototype. A new item
is categorized as an instance of the concept if it is sufficiently similar
to this prototype. The prototype consists of a set of features $\{a_1, a_2,
a_3...
a_M\}$, with associated
weights or applicability values $\{x_1, x_2, x_3... x_M\}$, where $M$ is
the number of features
considered. The distance between a new item and the prototype can be
calculated as follows, where
$s$ indexes the test stimulus, $x_{sm}$ refers to applicability of $m^{th}$
feature to the
stimulus $s$, and $x_{pm}$ refers to applicability of $m^{th}$ feature to
the prototype:

\be \label{eq:prototype}
d_s = \sqrt{\sum_{m=1}^M(x_{sm}-x_{pm})^2}
\ee
The smaller the value of $d_s$ for a given item, the more representative it
is of the concept.
Thus concept membership is graded, a matter of degree.

According to the {\it exemplar theory}, ({\it e.g.}
Medin, Altom, and
Murphy 1984, Nosofsky 1988, 1992, Heit and Barsalou 1996) a concept is
represented by, not defining
or characteristic
features, but a set of {\it instances} of it stored in memory.
Thus each of the $\{E_1, E_2, E_3,... E_N\}$ exemplars has a set $\{a_1,
a_2, a_3,... a_M\}$
of features with associated weights $\{x_1, x_2, x_3,... x_M\}$. A new
item is categorized
as an instance of concept if it is sufficiently similar to one or more of
these previously
encountered instances. For example, Storms {\it et al.} (2000) used the
following distance
function, where $s$ indexes the test stimulus, $x_{sm}$ refers to
applicability of $m^{th}$
feature to stimulus $s$, and $x_{nm}$ refers to applicability of $m^{th}$
feature to $n^{th}$
most frequently generated exemplar:

\be \label{eq:exemplar}
d_s = \sum_{n=1}^N\sqrt{\sum_{m=1}^M(x_{sm}-x_{nm})^2}
\ee
Once again, the smaller the value of $d_s$ for a given item, the more
representative it is of the concept.

Note that these theories have difficulty accounting for why items that
are dissimilar or even opposite might nevertheless belong together; for
example, why {\bf white} might be more likely to be categorized with {\bf
black} than with {\bf flat}, or why {\bf dwarf} might be more likely to be
categorized with {\bf giant} than with, say, {\bf salesman}. The only way
out is
to give the set of relevant variables or measurements or contexts the same
status as the {\it values} for
those variables or measurements or contexts {\it i.e.} to lump together as
features not only things like
`large' but also things like `has a size' or `degree to which size is
relevant'.

According to another approach to concepts, referred to as the {\it
theory theory}, concepts take the form of `mini-theories' ({\it e.g.}
Murphy and Medin 1985) or schemata (Rummelhart and Norman 1988), in which
the causal relationships amongst features or properties are identified.
A mini-theory contains knowledge concerning both which variables or
measurements are relevant, and the values obtained for them.
This does seem
to be a step toward a richer understanding of concept representation,
though many limitations have been pointed out (see for example Komatsu
1992, Fodor 1994, Rips 1995). Clearly, the calculation of conceptual
distance is less straightforward, though to us this reveals not so much a
shortcoming of the theory theory, but of the concept of conceptual distance
itself. In our view, concepts are not distant from one another at all, but
interwoven, and this interwoven structure cannot be observed directly, but
only indirectly, as context-specific instantiations. For example, the
concept {\bf egg} will be close to {\bf sun} in the
context `sunny side up' but far in the context `scrambled', and in the
context of the Dr. Suess book `Green Eggs and Ham' it acquires the feature
`green'.

Yet another theory of concepts, which captures their mutable,
context-dependent nature, but at the cost of increased vagueness, is {\it
psychological essentialism}. The basic idea is that instances of a concept
share a hidden essence which defines its true nature ({\it e.g.} Medin and
Ortony 1989). In this paper we attempt to get at this notion in a more
rigorous and explicit way than has been done.

\subsection{Membership Assessments for Conjunctive Categories}
The limitations of representational theories became increasingly
evident through experiments involving conjunctions of concepts. One such
anomalous phenomenon is the so-called {\it guppy effect}, where a guppy
is {\it not} rated as a good example of the concept {\bf pet}, nor of the
concept {\bf fish}, but it {\it is} rated as a good example of {\bf pet
fish} (Osherson and Smith 1981)\footnote{In fact, it has been demonstrated
experimentally that other conjunctions are better examples of the `guppy
effect' than pet fish
(Storms {\it et al.} 1998), but since this example is well-known, we will
continue to use it here.}. Representational theories cannot account for
this. Using the prototype approach, since a guppy is neither a typical pet
nor a typical fish, $d_s$ for the guppy stimulus is large for both {\bf
pet} and {\bf fish}, which is difficult to reconcile with the empirical
result that it is small for {\bf pet fish}. Using the exemplar approach,
although a guppy is an exemplar of both {\bf pet} and {\bf fish}, it is
unlikely to be amongst the $n$ most frequently generated ones. Thus once
again $d_s$ is large for both {\bf pet} and {\bf fish}, which is difficult
to reconcile with it being small for {\bf pet fish}.

The problem is not solved using techniques from fuzzy set mathematics such as
the {\it minimum rule model}, where the typicality of a
conjunction (conjunction typicality) equals the minimum of
the typicalities of the two constituent concepts (Zadeh 1965, 1982). (For
example, the typicality rating for {\bf pet fish} certainly does
not equal the minimum of that for {\bf pet} or {\bf fish}.)
Storms {\it et al.}
(2000) showed that a weighted and calibrated version of the minimum rule
model can
account for a substantial
proportion of the variance in typicality ratings for conjunctions
exhibiting the guppy effect.
They suggested the effect could be due to the existence of {\it contrast
categories}, the idea
being that a concept such as {\bf fruit} contains not
only information
about fruit, but information about categories that are related to, yet
different from, fruit.
Thus, a particular item might be a better exemplar of the concept {\bf
fruit} if it not only has
many features in common with exemplars of {\bf fruit} but also few features
in common with
exemplars of {\bf vegetables} (Rosch and Mervis 1975). However, another
study provided negative
evidence for contrast categories (Verbeemen {\it et al.} in press).

Nor does the theory theory or essence approach get us closer to solving the
conjunction problem. As Hampton (1997) points out, it is
not clear how a set of syntactic rules for combining or interpreting
combinations of mini-theories could be
formulated.

\subsection{`Emergence' and Loss of Properties During Conjunction}
An even more perplexing problem facing theories of concepts is that, as
many studies ({\it e.g.} Hastie {\it et. al.} 1990, Kunda
{\it et. al.} 1990, Hampton 1997)
have shown, a conjunction often possesses features which are said to be
{\it emergent}: not true of its' constituents. For example, the properties
`lives in cage' and `talks' are considered true of {\bf pet birds}, but not
true of {\bf pets} or {\bf birds}.

Representational theories are not only incapable of {\it predicting} what
sorts of features will emerge (or disappear) in the
conjunctive concept, but they do not even provide a place in the formalism
for the gain (or loss) of features. This problem stems back to a limitation
of the mathematics underlying not only
representational theories of concepts (as well as compositional theories of
language) but also classical physical theories. The
mathematics of classical physics only allows one to describe a
composite or joint entity by means of the product state
space of the state spaces of the two subentities. Thus if $X_1$ is the
state space of the first subentity, and $X_2$ the state
space of the second, the state space of the joint entity is the Cartesian
product space $X_1 \times X_2$. For this reason,
classical physical theories cannot describe the situation wherein two
entities generate a new entity with properties not strictly
inherited from its constituents.

One could try to solve the problem {\it ad hoc} by starting all over again
with a
new state space each time there appears a state that was not possible given
the previous state
space; for instance, every time a conjunction like {\bf pet bird} comes
into existence. However, this happens every time one generates a sentence
that has not been used before, or even uses the same sentence in a slightly
different context. Another
possibility would be to make the state space infinitely large to begin
with. However, since we hold
only a small number of items in mind at any one time, this is not a viable
solution to the problem
of describing what happens in cognition. This problem is hinted at by Boden
(1990), who uses
the term {\it impossibilist creativity} to refer to creative acts that not
only {\it explore} the existing
state space but {\it transform} that state space; in other words, it
involves the spontaneous
generation of new states with new properties.

\subsection{The `Obligatory Peeking' Principle}
In response to difficulties concerning the transformation of concepts, and
how mini-theories combine to form conjunctions,
Osherson and Smith (1981) suggested that, in addition to a modifiable
mini-theory, concepts have a stable definitional
{\it core}. It is the core, they claim, that takes part in the combining
process. However, the notion of a core does not
straightforwardly solve the conjunction problem. Hampton (1997) suggests
that the source of the difficulty is that in
situations where new properties emerge during concept conjunction, one is
making use of world knowledge, or `extensional
feedback'. He states: `We can not expect any model of conceptual
combination to account directly for such effects, as they
clearly relate to information that is obtained from another source---namely
familiarity with the class of objects in the
world' (p. 148). Rips (1995) refers to this as the {\it No Peeking
Principle}. Rips' own version of a dual theory
distinguishes between representations-of and representations-about, both of
which are said to play a role in
conjunction. However, he does not claim to have solved the problem
of how to describe concepts and their
conjunctions, noting `It seems likely that part of the semantic story will
have to include external causal connections that run through the referents
and their representations' (p. 84).

Goldstone and Rogosky's (in press) ABSURDIST algorithm is a move in this
direction. Concept meaning depends on a web of relations to other concepts
in the same domain, and the algorithm uses within-domain similarity
relations to translate across domains. In our contextualized approach, we
take this even further by incorporating not just pre-identified relations
amongst concepts, but new relations made apparent in the context of a
particular stimulus situation, i.e. the external world. We agree that it
may be beyond
our reach to predict exactly how world knowledge will come into play in
every particular case. However, it is at least possible to put forth a
theory of concepts that not only {\it allows} `peeking',
but in a natural (as opposed to {\it ad hoc}) way provides a place for it.
In fact, in our model, peeking (from either another concept, or an external
stimulus) is obligatory; concepts require a peek, a context, to actualize
them in some form (even if it is just the most prototypical
form). The core or essence of a concept is viewed as a source of potentiality
which requires some context to be dynamically actualized, and which thus
cannot be described in a context-independent manner (except as a
superposition of every possible context-driven instantiation of it). In
this view, each of the two concepts in a conjunction constitutes a context
for the other that `slices through' it at a particular angle, thereby
mutually actualizing one another's potentiality in a specific way. As a
metaphorical explanatory aid, if concepts were apples, and the stimulus a
knife, then the qualities of the knife would determine not just which apple
to slice, but which direction to slice through
it. Changing the knife (the context) would expose a different face of the
apple (elicit a different version of the
concept). And if the knife were to slash through several apples (concepts)
at once, we might end up with a new kind of apple (a conjunction).

\section{Two Cognitive Modes: Analytic and Associative}
We have seen that, despite considerable success when limited to simple
concepts like {\bf bird}, representational theories run into trouble when
it comes to conjunctions like {\bf pet bird} or even {\bf green bird}. In
this section we address the question: why would they be so good for
modeling many aspects of cognition, yet so poor for others?

\subsection{Creativity and Flat Associative Hierarchies} \label{sec:creatflat}
It is widely suggested that there exist two forms of thought ({\it e.g.}
James 1890, Piaget 1926, Neisser 1963, Johnson-Laird 1983, Dennett 1987,
Dartnell 1993, Sloman 1996, Rips 2001). One is a focused, evaluative {\it
analytic mode}, conducive to analyzing relationships of {\it cause and
effect}. The other is an intuitive, creative {\it associative mode} that
provides access to remote or subtle connections between features that may
be {\it correlated} but not necessarily causally related. We suggest that
while representational theories are fairly adequate for predicting and
describing the results of cognitive processes that occur in the analytic
mode, their shortcomings are revealed when it comes to predicting and
describing the results of cognitive processes that occur in the associative
mode, due to the more contextual nature of cognitive processes in this mode.

Since the associative mode is thought to be more evident in creative
individuals, it is useful at this point to look briefly at some of the
psychological attributes associated with creativity. Martindale (1999) has
identified a cluster of such attributes, including defocused attention
(Dewing and Battye 1971, Dykes and McGhie 1976, Mendelsohn 1976), and high
sensitivity (Martindale and Armstrong 1974, Martindale 1977), including
sensitivity to subliminal impressions; that is, stimuli that are perceived
but of which we are not {\it conscious} of having perceived (Smith and Van
de Meer 1994).

Another characteristic of creative individuals is that they have {\it flat
associative hierarchies} (Mednick 1962). The steepness of an individual's
associative hierarchy is measured experimentally by comparing the number of
words that individual generates in response to stimulus words on a word
association test. Those who generate only a few words in response to the
stimulus have a {\it steep} associative hierarchy, whereas those who
generate many have a {\it flat} associative hierarchy. Thus, once such an
individual has run out of the more usual associations ({\it e.g.} {\bf
chair} in response to {\bf table}), unusual ones ({\it e.g.} {\bf elbow} in
response to {\bf table}) come to mind.

It seems reasonable that in a state of defocused attention and heightened
sensitivity, more features of the stimulus situation or concept under
consideration get processed. (In other words, the greater the value for $M$
in equations (\ref{eq:prototype}) and (\ref{eq:exemplar}) for prototype and
exemplar theories.) It also seems reasonable that flat associative
hierarchies result from memories and concepts being more richly etched
into memory; thus there is a greater likelihood of an associative link
between any two concepts. The experimental evidence that flat associative
hierarchies are associated with defocused attention and heightened
sensitivity suggests that the more features processed, the greater the
potential for associations amongst stored
memories and concepts. We can refer to the detail with which items are
stored in memory as associative richness.

\subsection{Activation of Conceptual Space: Spiky versus Flat}
We now ask: how might different individuals, or a single individual under
different circumstances, vary with respect to
degree of detail with which the stimulus or object of thought gets etched
into memory, and resultant degree of
associative richness\footnote{In (Gabora 2000, 2002a, 2002b) the cognitive
mechanisms underlying creativity are
discussed in greater detail.}? A means of accomplishing this can be seen in
neural networks that use a radial basis
function (RBF), where each input activates a hypersphere of hidden nodes,
with activation tapering off in all directions
according to a (usually) Gaussian distribution of width
$\sigma$ (Willshaw and Dayan, 1990, Hancock {\it et al.}, 1991, Holden and
Niranjan, 1997, Lu {\it et al.}
1997)\footnote{This enables one part of the network to be modified without
interfering with the capacity of other parts
to store other patterns.}. Thus if $\sigma$ is small, the input activates
few memory
locations but these few are hit hard; we say the activation function is
{\it spiky}. If $\sigma$ is large, the input activates many memory
locations to an almost equal degree; we say the activation function is
relatively {\it flat}.

Whether or not human memory works like a RBF neural network, the idea
underlying them suggests a basis for the distinction between associative
and analytic modes of thought. We will use the terms spiky and flat
activation function to refer to the extent to which memory gets activated
by the stimuli or concepts present in a given cognitive state, bearing in
mind that this may work differently in human cognition than in a neural
network.\footnote{In a neural network, the center of the RBF (as well as
the value for $\sigma$) are determined during a training phase. However,
this is not necessary if memory locations simply differ in their capacity to
detect and respond to different features. According to the
temporal coding hypothesis, different features or stimulus dimensions are
carried by different frequencies like a radio broadcast system, and each
memory location is attuned to respond to a slightly different frequency, or
set of frequencies. ({\it e.g.} Stumpf 1965, Campbell and Robson 1968,
Perkell and Bullock 1968, Emmers 1981, Lestienne and Strehler 1987, Abeles
{\it et al.} 1993, Mountcastle 1993, Cariani 1995, 1997, Metzinger 1995,
Lestienne 1996, Riecke and Warland 1997, for reviews see De Valois and De
Valois 1988, Pribram 1991). As Cariani points out, temporal coding
drastically simplifies the problem of how the brain coordinates, binds, and
integrates information. The greater the number of stimulus frequencies
impacting the memory architecture, the greater the number of memory
locations that respond.} The basic idea then is that when the activation
function is spiky, only the most typical, central features of a stimulus or
concept are processed. This is conducive to analytic thought where
remote associations would be merely a distraction; one does not want to get
sidetracked by features that are
atypical, or modal (Rips 2001), which appear only in imagined or
counterfactual instances. However, as the
number of features or stimulus dimensions increases, features that are less
central to the concept that best
categorizes it start to get included, and these features may in fact make
it defy straightforward
classification as strictly an instance of one concept or another. When the
activation function is relatively
flat, more features are attended and participate in the process of
activating and evoking from memory;
atypical as well as typical ones. Therefore, more memory locations
participate in the release of
`ingredients' for the next instant. These locations will have previously
been modified by (and can therefore
be said to `store' in a distributed manner) not only concepts that
obviously share properties with the
stimulus, but also concepts that are correlated with it in unexpected ways.
A flat activation function is
conducive to creative, associative thought because it provides a high
probability of evoking one or more
concepts not usually associated with the stimulus.

Thus we propose that representational theories---in which concepts are
depicted as fixed sets of attributes---are adequate for modeling analytical
processes, which establish relationships of cause and effect amongst
concepts in their most prototypical forms. However they are not adequate
for modeling associative processes, which involve the identification of
correlations amongst more richly detailed, context-specific forms of
concepts. In a particular instant of thought or
cognitive state in the associative mode, aspects of a situation the
relevance of which may not be readily apparent,
or relations to other concepts which have gone unnoticed---perhaps of an
analogical or metaphorical nature---can `peek through'. A cognitive state
in which a new relationship amongst concepts is identified is a state of
{\it potentiality}, in the sense that the newly identified relationship
could be resolved different ways depending on the contexts one encounters,
both immediately, and down the road. For example, consider the cognitive
state of the person who thought up the idea of building a snowman. It seems
reasonable that this involved thinking of {\bf snow} not just in terms of
its most typical features such as `cold' and `white', but also the less
typical feature `moldable'. At the instant of inventing {\bf snowman} there
were many ways of resolving how to give it a nose.
However, perhaps because the inventor happened to have a carrot handy, the
concept {\bf snowman} has come to acquire the feature `carrot nose'.

\section{A Formalism that Incorporates Context}
We have seen that representational theories are good for describing and predicting the outcomes of experiments involving single concepts, particularly typical contexts, or relationships of {\it causation}. But they are not good at predicting the outcomes of experiments involving concepts in atypical contexts, or creatively blended together through the discovery of some unlikely {\it correlation}. This story has a precedent. Classical physics does exceedingly well at
describing and predicting relationships of causation, but it is much
less powerful in dealing with results of experiments that entail
sophisticated
relationships of correlation. It cannot describe certain types of
correlations that appear when quantum entities interact and combine to form
joint entities. According to
the dynamical evolution described by the Schr\"odinger equation, whenever
there is interaction between quantum entities, they spontaneously
enter an entangled state that contains new properties that the original
entities
did not have. The description of this birth of new states and new properties
required
the quantum mechanical formalism.

Another way in which the shortcomings of classical mechanics were
revealed had to do in a certain sense with the issue of `peeking'. A
quantum particle could not be observed without disturbing it; that is,
without changing its state. Classical mechanics could describe situations
where the effect of a measurement was negligible, but not situations where
the measurement intrinsically influenced the evolution of the entity. The best
it could do is to avoid as much as possible any
influence of the measurement on the physical entity under study. As a
consequence, it had to limit its set of valuable
experiments to those that have almost no effect on the physical entity
(called observations).
It could not incorporate the context generated by a measurement directly
into the formal description of the physical entity. This too required
the quantum formalism.

In this section we first describe the pure quantum formalism. Then we briefly
describe the generalization of it that we apply to the
description of concepts.

\subsection{Pure Quantum Formalism}

In quantum mechanics, the state of a physical entity can change in two
ways: (1) under the influence of a measurement context, and
this type of change is called {\it collapse}, and (2) under the influence
of the environment as a whole, and this change is called
{\it evolution}. In quantum mechanics, a
state $\psi$ is represented by a unit vector of a
complex Hilbert space ${\cal H}$, which is a vector space over the complex
numbers equipped with an inproduct (see Appendix
\ref{appendix01}). A property of the quantum entity is described by a
closed subspace of the complex Hilbert space or by the
orthogonal projection operator $P$ corresponding to this closed subspace,
and a measurement context by a self-adjoint operator on
the Hilbert space, or by the set of orthogonal projection operators that
constitute the spectral family of this self-adjoint
operator (see appendix
\ref{appendix02}). If a quantum entity is in
a state $\psi$, and a measurement context
is applied to it, the state $\psi$ changes to the
state
\be
{P(\psi) \over \|P(\psi)\|}
\ee
where $P$ is the projector of the spectral family of the
self-adjoint operator corresponding to the outcome of the
measurement. This change of state is more specifically what is meant by the
term collapse. It is a probabilistic change and the
probability for state $\psi$ to change to state ${P(\psi) /
\|P(\psi)\|}$ under influence of the measurement context is
given by
\be
\langle \psi, P(\psi) \rangle
\ee
where $\langle\ ,\ \rangle$ is the inproduct of the Hilbert space (see
appendix \ref{appendix02}).

The state prior to, and independent of, the measurement, can be
retrieved as a theoretical object---the unit vector of complex
Hilbert space---that reacts to all possible measurement contexts in
correspondence with experimental results. One of the merits of quantum
mechanics is that it made it possible to describe the undisturbed and
unaffected state of an entity even if most
of the experiments needed to measure properties of this entity
disturb this state profoundly (and often even destroy it). In
other words, the message of quantum mechanics is that is possible to
describe a reality that only can be known through acts that alter this reality.

There is a distinction in quantum mechanics between similarity in terms
of which measurements or contexts are relevant, and similarity in terms of
values for these measurements (a distinction which we saw in section two
has not been present in theories of concepts). Properties for which the same
measurement---such as the measurement of spin---is relevant are said to be
{\it compatible} with respect to this measurement. One of the axioms of
quantum mechanics---called {\it weak modularity}---is the requirement that
orthogonal properties---such as 'spin up' and 'spin down'---are
compatible.

In quantum mechanics, the conjunction problem is seriously addressed, and to
some extent solved, as follows. When quantum entities
combine, they do not stay separate as classical physical entities tend to
do, but enter a state of {\it entanglement}. If ${\cal H}_1$ is the Hilbert
space
describing a first subentity, and ${\cal H}_2$ the Hilbert space describing
a second subentity, then the joint entity is
described in the tensor product space ${\cal H}_1 \otimes {\cal H}_2$ of the
two Hilbert spaces ${\cal H}_1$ and ${\cal H}_2$. The tensor product always
allows for the emergence of new
states---specifically the entangled states---with new properties.

The presence of entanglement---{\it i.e.} quantum structure---can be tested
for by determining whether correlation experiments on
the joint entity violate Bell inequalities (Bell 1964). Pitowsky (1989)
proved that if Bell inequalities are satisfied for a set
of probabilities concerning the outcomes of the considered experiments,
there exists a classical Kolmogorovian probability
model that describes these probabilities. The probability can then be
explained as being due to a lack of
knowledge about the precise state of the system. If,
however, Bell inequalities are violated, Pitowsky proved that no such
classical Kolmogorovian probability model exists. Hence,
the violation of Bell inequalities shows that the probabilities involved
are nonclassical. The only type of nonclassical
probabilities that are well known in nature are the quantum probabilities.

\subsection{Generalized Quantum Formalism}
The standard quantum formalism has been generalized, making it possible to
describe changes of
state of entities with any degree of contextuality, whose structure
is not purely classical
nor purely quantum, but something in between (Mackey 1963, Jauch 1968,
Piron 1976, 1989, 1990, Randall and Foulis 1976,
1978, Foulis and Randall 1981, Foulis, Piron, and Randall 1983, Pitowsky
1989, Aerts 1993, 2002, Aerts and Durt
1994a,b). The generalizations of
the standard quantum formalism we use as the core mathematical
structure to replace the Hilbert space of standard quantum mechanics has the
structure of a {\it lattice}, which represents the set of 
properties of the entity under consideration. Many different types
of lattices have been introduced, depending on the type of
generalized approach and on the particular problem under study. This has
resulted in mathematical structures that are more
elaborate than the original lattice structure, and it is one of them,
namely the {\it state context property system}, or $SCOP$, that we
take as a starting point here.

Let us now outline the basic mathematical structure of a $SCOP$. It
consists of three sets and two functions, denoted:
\be
(\Sigma, {\cal M}, {\cal L}, \mu, \nu)
\ee
where:
\begin{itemize}
\item $\Sigma$ is the set of possible states.
\item ${\cal M}$ is the set of relevant contexts.
\item ${\cal L}$ is the lattice which describes the relational structure of
the set of relevant properties or features.
\item $\mu$ is a probability function that describes how a couple $(e, p)$,
where $p$ is a state, and $e$ a context, transforms to a
couple $(f, q)$, where $q$ is the new state (collapsed state for
context $e$), and $f$ the new context.
\item $\nu$ is the weight or applicability of a certain property, given a
specific
state and context.
\end{itemize}
The structure ${\cal L}$ is that of a complete, orthocomplemented lattice.
This means that

\begin{itemize}
\item A partial order relation denoted $<$ on ${\cal L}$ representing that
the implication of properties, {\it i.e.} actualization of one
property implies the actualization of another.
For $a, b \in {\cal L}$ we have
\bea
a < b \Leftrightarrow \ {\rm if}\ a\ {\rm then}\ b\ {\rm}
\eea

\item Completeness: infimum (representing the conjunction and denoted
$\wedge$) and supremum
(representing the disjunction and denoted $\vee$) exists for any subset of
properties. $0$, minimum
element, is the infimum of all elements of ${\cal L}$ and $I$, maximal
element, is the supremum of all
elements of
${\cal L}$.

\item Orthocomplemented: an operation $^\perp$ exists, such that for $a, b \in
{\cal L}$ we have
\bea
(a^\perp)^\perp &=& a \\
a < b &\Rightarrow& b^\perp < a^\perp \\
a \wedge a^\perp = 0&,& a \vee a^\perp = I
\eea
Thus $a^\perp$ is the `negation' of $a$.

\item Elements of ${\cal L}$ are weighted. Thus for state $p$, context $e$
and property $a$ there exists weight
$\nu(p, e, a)$, and for $a \in {\cal L}$
\be
\nu(p, e, a) + \nu(p, e, a^\perp) = 1
\ee
\end{itemize}
These general
formalisms describe much more than is needed for quantum mechanics, and in
fact, standard quantum
mechanics and classical
mechanics fall out as special cases (Aerts 1983). For the $SCOP$
description of a pure quantum entity, see Appendix \ref{appendix03}.

It is gradually being realized that the generalized quantum formalisms have
relevance to the macroscopic world ({\it e.g.} Aerts 1991, Aerts {\it et
al.} 2000a,b). Their
application beyond the domain that originally gave
birth to them is not as strange as it may seem. It can even be viewed as an
unavoidable sort of
evolution, analogous to what has been observed for chaos and complexity
theory. Although chaos
and complexity theory were developed for application in inorganic physical
systems, they quickly
found applications in the social and life sciences, and are now thought of
as domain-general
mathematical tools with broad applicability. The same is potentially true
of the mathematics
underlying the generalized quantum formalisms. Although originally
developed to describe the
behavior of entities in the microworld, there is no reason why their
application should be
limited to this realm. In fact, given the presence of potentiality and
contextuality in
cognition, it seems natural to look to these formalisms for guidance in the
development of a
formal description of cognitive dynamics.

\section{Application of SCOP to Concepts}
In this section we apply the generalized quantum formalism---specifically
the $SCOP$---to cognition, and show what concepts reveal themselves to be
within this framework. To do this we must make a number of subtle but
essential points. Each of these points may appear strange and not
completely motivated in itself, but together they deliver a clear and
consistent picture of what concepts are.

We begin by outlining
some previous work in this direction. Next we present the mathematical
framework. Then we examine more closely the roles of potentiality, context,
collapse, and actualization. Finally we will focus more specifically on how
the formalism is used to give a measure of conceptual distance. This is
followed up on in the next section which shows using a specific example how
the formalism is applied to concept conjunction.

\subsection{Previous Work}
One of the first applications of these generalized formalisms to cognition
was modeling the decision making process. Aerts and Aerts (1994) proved
that in situations where one moves from a state of indecision to a decided
state (or {\it vice versa}), and the change of state is context-dependent,
the probability distribution necessary to describe it is non-Kolmogorovian.
Therefore a classical probability model cannot be used. Moreover, they
proved that such situations {\it can} be accurately described using these
generalized quantum mathematical formalisms. Their mathematical treatment
also applies to the situation where a cognitive state changes in a
context-dependent way to an increasingly specified
conceptualization of a stimulus or situation. Once again, context induces a
nondeterministic change of the cognitive state which
introduces a non-Kolmogorovian probability on the state space. Thus, a
nonclassical (quantum or generalized quantum) formalism is
necessary.

Using an example involving the concept {\bf cat} and instances of cats, we
proved that Bell inequalities are violated in the
relationship between a concept and specific instances of it (Aerts {\it et
al.} 2000a). Thus we have evidence that this formalism
reflects the underlying structure of concepts. In (Aerts {\it et al.}
2000c) we show that this result is obtained because of the presence of nonlocality, that is, EPR-type correlations, amongst the properties of concepts. The
EPR nature of these correlations arises because of how concepts
exist in states of potentiality, with the presence or absence of particular
properties being determined {\it in the process of} the evoking
or actualizing of the concept. In such situations, the mind handles
disjunction in a quantum manner.  It is to be expected that such
correlations exist not only amongst different instances of a single
concept, but amongst different related concepts, which makes the notion
of conceptual distance even more suspect.

\subsection{Mathematical Framework}

In the development of this approach, it became clear that to be able to
describe contextual interactions and conjunctions of concepts, it is useful
to think not just in terms of concepts {\it per se}, but in terms of
the cognitive states that instantiate them. Each concept is potentially
instantiated by many cognitive states; in other words, many thoughts or
experiences
are interpreted in terms of any
given concept. This is why we first present the mathematical structure that
describes an entire conceptual system, or
mind. We will then illustrate how concepts appear in this structure.  We
use the mathematical structure of a state
context property system or $SCOP$:
\be
(\Sigma, {\cal M}, {\cal L}, \mu, \nu)
\ee
where
\begin{itemize}

\item $\Sigma$ is the set of all possible cognitive states, sometimes
referred to as conceptual space. We use symbols $p, q, r,
\dots$ to denote states.
\item ${\cal M}$ is the set of relevant contexts that can influence how a
cognitive state is categorized or conceptualized. We use symbols $e, f, g,
\ldots$ to denote contexts.
\item ${\cal L}$ is the lattice which describes the relational structure of
the set of relevant properties or features. We use symbols $a, b, c,
\ldots$ to denote features or properties.
\item $\mu$ is a probability function that describes how a couple $(e, p)$,
where $p$ is a state, and $e$ a context, transforms
to a couple $(f, q)$, where $q$ is the new state (collapsed state for
context $e$), and $f$ the new context.
\item $\nu$ is the weight or
applicability of a certain property, given a specific state and context.
\end{itemize}
By cognitive states we mean states of the mind (the mind being the entity
that experiences them). Whereas the
two sets $\Sigma$ and ${\cal M}$, along with the function
$\mu$, constitute the set of possible cognitive states and the contexts
that evoke them, the set ${\cal L}$ and the function $\nu$, describe
properties of these states, and their weights. In general, a cognitive
state $p \in \Sigma$ under context $e$ (the stimulus) changes to state $q
\in \Sigma$ according to probability function $\mu$. Even if the stimulus
situation itself does not change, the change of state from $p$ to $q$
changes the context ({\it i.e.} the
stimulus is now experienced in the context of having influenced the change
of state from $p$ and $q$). Thus we have a new context $f$. For a more
detailed exposition of $SCOP$ applied to cognition, see
appendix \ref{appendix04}.

\subsection{How Concepts Appear in the Formalism}
We denote concepts by the symbols $A, B, C, \ldots$, and the set of all
concepts ${\cal A}$. A concept appears in the formalism as a subentity of
this entire cognitive system, the
mind\footnote{To exactly express mathematically how a concept constitutes a
subentity of the mind we must define what
are the morphisms of the $SCOP$, and more generally the mathematical
category ${\bf SCOP}$ of the state context
property systems and their morphisms. We use the word category here as it
is used in the mathematical discipline {\it category theory}. For details
see (Aerts 2002, Aerts
and Gabora 2002) where the category ${\bf SCOP}$, and the notion of
subentity, are discussed at length.}.  This means
that if we consider a cognitive state $p \in \Sigma$, for each concept $A
\in {\cal A}$, there exists a corresponding
state $p_A$ of this concept. The concept $A \in {\cal A}$ is described by
its own $SCOP$ denoted $(\Sigma_A, {\cal M},
\mu_A, {\cal L}_A, \nu_A)$, where $\Sigma_A$ is the set of states of this
concept, and ${\cal M}$ is the set of
contexts.  Remark that ${\cal M}$ is the same for different concepts, and
for the mind as a whole, because all contexts
that are relevant for the mind as a whole are also relevant for a single
concept. Furthermore,
$\mu_A$ describes the probabilities of collapse between states and contexts
for this concept, and ${\cal L}_A$ and $\nu_A$ refer to the set of features
and weights relevant to concept $A$. When we speak of the potentiality of a
concept, we refer to the totality of ways in which it could be actualized,
articulated, or experienced in a cognitive state, given all the different
contexts in which it could be relevant.

\subsubsection{Instantiation of Concept Actualizes Potential}
For a set of concepts $\{A_1, A_2, \ldots, A_n, \ldots\}$, where $A_i \in
{\cal A}\ \forall i$, the cognitive state $p$
can be written $(p_{A_1}, p_{A_2}, p_{A_3}, \ldots, p_{A_n}, \ldots\}$,
where each $p_{A_i}$ is a state of concept
$A_i$. For a given context $e \in {\cal M}$, each of these states $p_A$
could be a potentiality state or a collapsed
state. Let us consider the specific situation where the cognitive
state $p$ instantiates concept $A_m$. What this explicitly means
is that $p_{A_m}$,
the state of
concept $A_m$, becomes an actualized cognitive state, and this corresponds
to the evoking of concept $A_m$. At the
instant $A_m$ is evoked in cognitive state $p$, its potentiality is
momentarily deflated or collapsed with
respect to the given context $e$.

\subsubsection{Uninstantiated Concepts Retain Potential}
Let us continue considering the specific situation where state $p$
instantiates concept $A_m$ under
context $e$. For each concept $A_i$ where $i \not=m$, no instantiation
takes place, and state
$p_{A_i}$ remains a complete potentiality state for context $e$. Thus,
concepts that are not evoked in the interpretation
of a stimulus to become present in the cognitive state retain their
potentiality. This means they are not limited to
a fixed set of features or relations amongst features. The formalism allows
for this because the state space where a
concept `lives' is not limited {\it a priori} to features thought to be
relevant. It is this that allows both their
contextual character to be expressed, with new features emerging under new
contexts. Given the right context were to
come along, any feature could potentially become incorporated into an
instantiation of it.

\subsubsection{Concepts as Contexts and Features}

In addition to appearing as subentities instantiated by cognitive states,
concepts appear in the formalism in two other ways. First, they can
constitute (part of) a context $e \in {\cal M}$. Second,
something that constitutes a feature or property $a \in {\cal L}$ in
one situation can constitute a concept in another; for instance, `blue' is
a property of the sky, but also one has a concept {\bf blue}.
Thus, the three sets $\Sigma$, ${\cal M}$ and ${\cal L}$, of a $SCOP$ are
all in some way affected by concepts.

\subsubsection{Conjunctions of Concepts}
As mentioned previously, the operation applied to pure quantum
entities is the tensor product. The algebraic operation we feel to be most
promising for the description of conjunction of concepts is the following.
In a $SCOP$, there is a straightforward connection between the state of the
entity under consideration at a
certain moment, and the set of properties that are actual at that moment
\footnote{In earlier versions of
$SCOP$, for example in (Aerts 1982, Piron 1976, 1989, 1990), both were even
identified.}. This makes it possible to, for a
certain fixed property $a \in {\cal L}$, introduce what is called the
relative $SCOP$ for $a$, denoted $(\Sigma, {\cal M},
\mu, {\cal L}, \nu)_a$. Suppose that $(\Sigma, {\cal M}, \mu, {\cal L},
\nu)$ describes concept $A$, then $(\Sigma, {\cal
M}, \mu, {\cal L}, \nu)_a$ describes concept $A$ given that property $a$ is
always actual for $A$. We could, for example,
describe with this structure the concept {\bf pet} where the property {\it
swims} is always actual. This would give us a
possible model for the conjunction of a noun concept with an adjective
concept. In the case of {\bf pet} and {\it swims}
this would come close to {\bf pet fish}, but of course, that this happens
is certainly not a general rule. For the case of
a conjunction of two nouns, if we want to try out the relative $SCOP$
construction, we would have to consider the
conjunctions of all possible features of the two nouns and derive from this
the $SCOP$ that would describe the conjunction
of the two nouns.

\subsection{Superposition, Potentiality Couples, and Change of Cognitive State}
We cannot specify with complete accuracy (1) the content of state $p$, nor
(2) the stimulus situation it faces, context $e$, nor (3) how the
two will interact. Therefore, any attempt to mathematically model the
transition from $p$ to $q$ must incorporate the possibility that the
situation could be interpreted in many
different ways, and thus many different concepts (or conjunctions of them)
being activated. Within
the formalism, it is the structure of the probability field $\mu$ that
describes this.
For a given state $p$, and another state $q \in \Sigma$ and contexts $e$
and $f \in {\cal M}$, the probability
$\mu(f, q, e, p)$ that state $p$ changes under the influence of context
$e$ to state $q$ (and that $e$ changes to $f$) will often be different from
zero. In the quantum language, we can express this by saying that
$p$ is a superposition state of all the states $q \in \Sigma$ such that the
probability $\mu(f, q, e, p)$ is nonzero for some $e, f \in {\cal M}$.
Note that whether or not $p$ is in a state of potentiality depends on the
context $e$. It is possible that state
$p$ would be a superposition state for $e$ but not for another context
$f$. Therefore, we use the term {\it potentiality couple} $(e,
p)$.

We stress that the potentiality couple is different from the
potentiality of a concept; the potentiality couple refers to the cognitive
state (in all its rich detail) with respect to a certain context
(also in all its rich detail), wherein a particular instantiation of some
concept (or conjunction of them) may be what is being subjectively
experienced. However, they are related in the sense that the potentiality
of $p$ decreases if concepts $A \in {\cal A}$ evoked in it enter
collapsed states.

\subsubsection{Collapse: Nondeterministic Change of Cognitive State}
Following the quantum terminology, we refer to the cognitive state
following the change of state under the influence of
a context as a {collapsed state}. Very often, though certainly not always,
a state $p$ is a superposition state with
respect to context $e$ and it collapses to state $q$ which is an
eigenstate\footnote{\noindent
The terminology `eigenstate' comes from quantum mechanics, and the word
`eigen' is the German for `proper'. This German terminology is
reminiscent of the German environment that gave birth to quantum
mechanics.} with respect to $e$, but a superposition state
with respect to the new context $f$. This is the case when couple $(e, p)$
refers to conception of stimulus prior to categorization, and
couple $(f, q)$ refers to the new situation after categorization has taken
place.

Recall that a quantum particle cannot be observed or `peeked at'
without disturbing it; that is, without inducing a change of state.
Similarly, we view concepts as existing in states of potentiality which
require a context---activation by a stimulus or other concept that
constitutes (perhaps partially) the present cognitive state---to be
elicited and thereby constitute the content (perhaps partially) of the next
cognitive state. However, just as in the quantum case, this `peeking'
causes the concept to collapse from a state of potentiality to a particular
context-driven instantiation of it. Thus, the stimulus situation plays the
role of the measurement by determining which
are the possible states that can be collapsed upon; it `tests' in some
way the potentiality of the associative network,
forces it to actualize, in small part, what it is capable of. A stimulus is
categorized as an instance of a specific concept
according to the extent to which the conceptualization or categorization of
it constitutes a context that collapses the
cognitive state to a thought or experience of the concept.

\subsubsection{Deterministic Change of Cognitive State}
A special case is when the couple $(e, p)$ is {\it not} a potentiality
couple. This means there exists a context $f$ and
a state
$q$, such that with certainty couple $(e, p)$ changes to couple $(f, q)$.
In this case we call $(e, p)$ a deterministic couple and $p$ a
deterministic state as a member of the couple $(e, p)$. An even more
special case is when the context $e$ does not provoke any change of the
state $p$. Then the couple $(e, p)$ is referred to as an eigencouple, and the
state $p$ an eigenstate as a member of the couple $(e, p)$.

\subsubsection{Retention of Potentiality during Collapse}
For a given stimulus $e$, the probability that the cognitive state $p$ will
collapse to a given concept $A$ is related to the algebraic structure of the
total state context property system $(\Sigma, {\cal M}, {\cal L}, \mu,
\nu)$, and most of all, to the probability field $\mu(f, q, e, p)$ that
describes how the stimulus and the cognitive state interact. It is clear
that, much as the potentiality of a concept (to be applicable in all sorts
of contexts) is reduced to a single actualized alternative when it
collapses to a specific instantiation, the potentiality of a stimulus (to
be interpreted in all sorts of ways) is diminished when it is interpreted
in terms of a particular concept. Thus, in the collapse process, the
stimulus loses potentiality. Consider as an example the situation that one
sees a flower, but if one were to examine it more closely, one would see
that it is a plastic flower. One possibility for how a situation such as
this gets categorized or conceptualized is that extraneous or modal
feature(s) are discarded, and the cognitive
state collapses completely to the concept that at first glance appears to best
describes it: in this case, {\bf flower}. We can denote this cognitive
state $p_1 \in \Sigma$. Some of the richness of the particular situation is
discarded, but what is gained is a straightforward way of framing it in
terms of what has come before, thus immediately providing a way to respond
to it: as one has responded to similar situations in the past. This is more
likely if one is in an
analytical mode and thus $\sigma$ is small, such that one does not encode
subtle details (such as `the flower is made of plastic').

However, a stimulus may be encoded in richer detail such that, in addition
to features known to be associated with the concept that could perhaps best
describe it, atypical or modal features are
encoded. This is more likely if one is in an associative mode, and thus
$\sigma$ is large. Let us denote as $p_2 \in \Sigma$ the state of
perceiving something that is flower-like, but that appears to be `made of
plastic'. The additional feature(s) of $p_2$ may make it more resistant to
immediate classification, thereby giving it potentiality. In the context of
wanting to make a room more cheerful it may serve the purpose of a flower,
and be treated as a flower, whereas in the context of a botany class it
will not. The state $p_2$, that retains potentiality may be close to
$p_1$, the completely collapsed state, but not identical to it.
In general, the flatter the activation function, the more features of the
stimulus situation are perceived and thus
reflected to and back from the associative network. Thus the more likely
that some aspects of the situation do not fall
cleanly into any particular category or concept, and therefore the more
potentiality present in the cognitive state, and
the more nonclassical the reflection process. Note that in an associative
mode, for a given cognitive state there will be
more features to be resolved, and so the variety of potential ways of
collapsing will tend to be greater. Hence the set of
states that can be collapsed to is larger.

\subsubsection{Loss of Potentiality through Repeated Collapse}
It seems reasonable that the presence of potentiality in a cognitive state
for a certain context is what induces the individual to continue thinking
about, re-categorizing, and reflecting on the stimulus situation. Hence
if the cognitive state is like $p_2$, and some
of the potentiality of the previous cognitive state was retained, this
retained potentiality can be collapsed in further
rounds of reflecting. Thus a stream of collapses ensues, and continues
until the stimulus or situation can be described in
terms of, not just one concept (such as {\bf flower}, but a complex
conjunction of concepts (such as `this flower is made
of plastic so it is not really a flower'). This is a third state $p_3$,
that again is a collapsed state, but of a more complex nature than the
first collapsed state $p_1$ was. But it is more stable with respect to the
stimulus than $p_1$ or $p_2$.

The process can continue, leading to a sequence of states $p_3, p_4, p_5,
\ldots$. With each iteration the
cognitive state changes
slightly, such that over time it may become possible
to fully interpret the stimulus situation in terms of it. Thus, the
situation eventually gets
interpreted as an instance of a new, more complex concept or category,
formed spontaneously through the conjunction of
previous concepts or categories during the process of reflection. The
process is contextual in that it is open to influence
by those features that did not fit the initial
categorization, and by new stimuli that happen to come along.

\subsection{Contextual Conceptual Distance}
We have claimed that for any concept, given the right context, any
feature could potentially become involved in its
collapse, and thus the notion of conceptual distance becomes less
meaningful. However, it is possible to obtain a measure of the
distance between states of concepts, potentiality states as well as
collapsed states (which can be
prototypes, exemplars, or imaginary constructions), and this is what the
formulas here measure.

\subsubsection{Probability Conceptual Distance} \label{sec:probabilitydistance}

First we define what we believe to be the most direct distance measure,
based on the probability field $\mu(f, q, e, p)$. This method is analogous
to the procedure used for calculating distance measure in quantum
mechanics. We first introduce a reduced probability:
\bea \label{eq:reducedprobability}
\mu: \Sigma \times {\cal M} \times \Sigma &\rightarrow& [0, 1] \\
(q, e, p) &\mapsto& \mu(q, e, p)
\eea
where
\be
\mu(q, e, p) = \sum_{f \in {\cal M}}\mu(f, q, e, p)
\ee
and $\mu(q, e, p)$ is the probability that state $p$ changes to state
$q$ under the influence of context $e$.

The calculation of probability conceptual distance is obtained using a
generalization of the distance in complex Hilbert space for the case of a
pure quantum situation, as follows:
\be \label{eq:quantumdistance}
d_\mu(q, e, p) = \sqrt{2(1-\sqrt{\mu(q, e, p)})}
\ee
We can also introduce the conceptual angle between two states, again making
use of the formula from pure quantum mechanics:
\be
\theta_\mu(q, e, p) = \arccos \mu(q, e, p)
\ee
We call $d_\mu$ the probability conceptual distance, or the $\mu$ distance,
and $\theta_\mu$ the probability conceptual angle, or the $\mu$ angle. For
details, see appendix \ref{appendix01} and equations
(\ref{eq:Hilbertdistance}) and (\ref{eq:Hilbertangle}), and remark that for
unit
vectors (\ref{eq:Hilbertdistance}) reduces to (\ref{eq:quantumdistance}).

Let us consider some special cases to see more clearly what is meant by
this distance and this angle. If $\mu(q, e, p) = 0$ we have $d_\mu(q, e, p)
= \sqrt{2}$ and $\theta_\mu(q, e, p) = {\pi \over 2}$. This corresponds to
the distance and angle between two orthogonal unit vectors in a
vectorspace. So orthogonality of states, when the probability that one
state changes to the other state is 0, represent the situation where the
distance is maximal ($\sqrt{2}$), and the angle is a straight angle (${\pi
\over 2}$). If $\mu(q, e, p) = 1$ we have $d_\mu(q, e, p) = 0$ and
$\theta_\mu(q, e, p) = 0$. This corresponds to the distance and angle
between two coinciding unit vectors in a vectorspace. So coincidence of
states---when the probability that one state changes to the other state =
1---represents the situation where the distance is minimal (0), and the
angle is minimal (0). For values of $\mu(q, e, p)$ strictly between $0$ and
$1$, we find a distance between $0$ and $\sqrt{2}$, and an angle between
$0$ and $\pi \over 2$.

It is important to remark that the distance $d_\mu(q, e, p)$ and angle
$\theta_\mu(q, e, p)$ between two states $p$ and $q$ is dependent on the
context $e$ that provokes the transition from $p$ to $q$. Even for a fixed
context, the distance does not necessarily satisfy the requirements that a
distance is usually required to satisfy in mathematics. For example, it is
not always the case that $d_\mu(q, e, p) = d_\mu(p, e, q)$, because the
probability $\mu(q, e, p)$ for $p$ to change to $q$ under context $e$ is
not necessarily equal to the probability $\mu(p, e, q)$ for $q$ to change
to $p$ under context $e$ \footnote{\noindent
It is easy to give an example that illustrates this. Consider a cognitive
state $p$ consisting of the concept {\bf bird}, and a context $e$ that
consists of the question `give me a feature of whatever concept you are
thinking of'. Consider another cognitive state $q$ consisting of the
concept {\bf feather}. The probability $\mu(q, e, p)$ will be reasonably
large. However, the probability $\mu(p, e, q)$, that a cognitive state for
{\bf feather} collapses to a cognitive state for {\bf bird} under the
context `give me a feature of feathers', will be close to zero, because
`bird' in not a feature of {\bf feather}.}.

\subsubsection{Property Conceptual Distance}
In order to illustrate explicitly the relationship between our approach and
the distance measures provided by the prototype and exemplar approaches
described previously, we define a second distance measure based on
properties. This distance measure
requires data on the probability of collapse of a cognitive state under the
influence of a context to a
cognitive state in which a particular feature is activated. In order to
define operationally
what this data refers to, we describe how it could be
obtained experimentally. One group of subjects is asked to consider one
particular concept
$A$, and this evokes in them cognitive state $p$. This state will be subtly
different for each subject, depending on the specific contexts which led
them to form these concepts, but there will be nevertheless commonalities.
A second group of subjects is asked to consider another concept $B$, which
evokes cognitive state $q$. Again, $q$ will be in some ways similar and in
some ways different for each of these subjects. The subjects are then asked
to give an example of `one' feature for each one of the considered
concepts. Thus, two contexts are at play: context $e$ that consists of
asking the subject to give a feature of the concept focused on in state
$p$, and context $f$ that consists of asking the subject to give a feature
of the concept focused on in state $q$. Thus we have two potentiality
couples $(e, p)$ and $(f, q)$. Suppose couple $(e, p)$ gives rise to the
list of features $\{b_1, b_2, \ldots, b_K\}$, and couple $(f, q)$ the list
of features $\{c_1, c_2, \ldots, c_L\}$. Some of the features may be
present on both lists, and others on only one. The two lists combined
generate a third list $\{a_1, a_2,\ldots, a_M\}$. Each feature $a_m$ is
active in a cognitive
state $r_m$ that one or more subjects collapses to under either context $e$
or $f$. By calculating the relative frequencies of these features, we
obtain an estimate of $\mu(r_m, e, p)$ and $\mu(r_m, f, q)$. The distance
between $p$ and $q$ is now defined as follows:

\be
d_p(q, e, f, p) = {\sqrt{2} \over \sqrt{M}}\sqrt{\sum_{m=1}^M(\mu(r_m, e,
p)-\mu(r_m, f, q))^2}
\ee
We call $d_p$ the probability property distance, or the $p$ distance, to
distinguish it
from $d_\mu$, the probability distance or $\mu$ distance.

Remark that to compare this distance $d_p$ to the $\mu$ distance $d_\mu$ we
introduce the renormalization factor ${\sqrt{2} /\sqrt{M}}$. This is to
make the maximal distance, which is attained if $|\mu(r_m, e, p)-\mu(r_m,
f, q)| = 1\ \forall\ m$, equal to $\sqrt{2}$.

We can also define a property conceptual distance based on weights of
properties. Given a set of features $\{a_1, a_2, \ldots,  a_M\}$, for each
of $p$ and
$q$, $\nu(p, e, a_m)$ is the weight of feature
$a_m$ for $p$ under context $e$, and $\nu(q, f, a_m)$ is the weight of
feature $a_m$ for $q$ under context $f$. The
distance between states $p$ and $q$ for the two concepts under contexts $e$
and $f$ respectively can be written as follows:
\be
d_w(q, e, f, p) = {\sqrt{2} \over \sqrt{M}}\sqrt{\sum_{m=1}^M(\nu(p, e,
a_m)-\nu(q, f, a_m))^2}
\ee
We call $d_w$ the weight property distance. It is clear that this distance
depends not only on $p$ and $q$, but also on the two contexts in which the
weights are obtained. How the weights depend on context follows partly from
the lattice ${\cal L}$, which describes the relational structure of the set
of features, and how this structure is related to the structure of the
probability field $\mu(f, q, e, p)$ which gives the probabilities of
collapse under a given context.

\subsubsection{Relationship Between the Two Distance Measurements}
\label{sec:weightprobability}
It would be interesting to know whether there is a relationship between the
distance measured using the probability field and the distance
measured using weighted properties. In pure quantum mechanics, these two
distances are equal (see appendix \ref{appendix03}, equations
(\ref{eq:weight}) and \ref{eq:collapseprobability})).

This could be tested experimentally as follows. Subjects are asked to give
a single feature of a given concept. We call $e$ the context
that consists of making this request. Since a concept $A$ evokes slightly
different cognitive states $p$ in different subjects, they do not
all respond with the same feature. Thus we obtain the set of features
$\{a_1, a_2, \ldots, a_M\}$. We denote the cognitive state of a given
subject corresponding to the naming of feature $a_m$ by $p_m$. The relative
frequency of feature $a_m$ gives us $\mu(p_m, e, p)$. In another
experiment, we consider the same concept $A$. We consider the
set of features $\{a_1, a_2, \ldots, a_M\}$ collected in the
previous experiment. Now subjects are asked to estimate the applicability
of these features to this concept. This gives us the weight
values $\nu(p, e, a_m)$. Comparing the values of $\mu(p_m, e, p)$  and
$\nu(p_m, e, p)$ makes it possible to find the relation between the two
distances $d_p$ and $d_a$.

\section{Application to the Pet Fish Problem}
We now present theoretical evidence of the utility of the contextual
approach using the Pet Fish Problem. Conjunctions such as this are dealt
with by incorporating context depen\-dency, as follows: (1) activation of
{\bf pet} still rarely causes activation of {\bf guppy}, and likewise (2)
activation of {\bf fish} still rarely causes activation of {\bf guppy}. But
now (3) {\bf pet fish} causes activation of the potentiality state {\bf
pet} {\it in the context of} {\bf pet fish} AND {\bf fish} {\it in the
context of} {\bf pet fish}. Since for this potentiality state, the
probability of collapsing onto the state {\bf guppy} is high, it is very
likely to be activated.

\subsection{The Probability Distance}

Let us now calculate the various distance measures introduced in the
previous section. We use equation (\ref{eq:quantumdistance}) for the
relevant states and contexts involved:

\be
d_\mu(q, e, p) = \sqrt{2(1-\sqrt{\mu(q, e, p)})}
\ee
where $\mu(q, e, p)$ is the probability that state $p$ changes to state $q$
under influence of context $e$. Two states and three contexts are at play
if we calculate the different
distances $d_\mu$ for the pet fish situation. State $p$ is the cognitive
state of a subject before any question is asked. Contexts $e$, $f$, and $g$
correspond to asking subjects to give an example of {\bf pet}, {\bf fish},
and {\bf pet fish} respectively. State $q$ corresponds to the cognitive
state consisting of the concept {\bf guppy}.

The transition probabilities are $\mu(q, e, p)$,
the probability that a subject answers `guppy' if asked to give an example
of {\bf pet}, $\mu(q, f, p)$, the probability that the subject answers
`guppy' if asked to give and example of {\bf fish}, and $\mu(q, g, p)$, the
probability that the subject answers `guppy' if asked to give and example
of {\bf pet fish}. The probability distances are then:
\bea
d_\mu(q, e, p) &=& \sqrt{2(1-\sqrt{\mu(q, e, p)})} \\
d_\mu(q, f, p) &=& \sqrt{2(1-\sqrt{\mu(q, f, p)})} \\
d_\mu(q, g, p) &=& \sqrt{2(1-\sqrt{\mu(q, g, p)})}
\eea
Since $\mu(q, e, p)$ and $\mu(q, f, p)$ are experimentally close to zero,
while $\mu(q, g, p)$ is close to 1, we have that $d_\mu(q, e, p)$ and
$d_\mu(q, f, p)$ are close to $\sqrt{2}$ (the maximal distance), and
$d_\mu(q, g, p)$ is close to zero.

\subsection{The Property Distances}
We only calculate explicitly the weight property distance $d_w$, since
this is the one calculated in representational approaches. The
probability property distance $d_p$is calculated
analogously.

Four states $p, q, r, s$ and four contexts $e, f, g, h$ are at play. The
states $p$, $q$, $r$, and $s$ are the cognitive states consisting of {\bf
guppy}, {\bf pet}, {\bf fish}, and {\bf pet fish} respectively. The
contexts $e$, $f$, $g$, $h$ are the experimental situations of being asked
to rate the typicality of {\bf guppy} as an instance of these four concepts
respectively. For an arbitrary feature $a_m$, the weights to consider are
$\nu(p, e, a_m)$, $\nu(q, f, a_m)$, $\nu(s, g, a_m)$ and $\nu(s, h, a_m)$.
The distances are:
\bea
d(p, e, f, q) &=& \sqrt{\sum_{m=1}^M(\nu(p, e, a_m)-\nu(q, f, a_m))^2} \\
d(p, e, g, r) &=& \sqrt{\sum_{m=1}^M(\nu(p, e, a_m)-\nu(r, g, a_m))^2} \\
d(p, e, h, s) &=& \sqrt{\sum_{m=1}^M(\nu(p, e, a_m)-\nu(s, h, a_m))^2}
\eea
Thus we have a formalism for describing concepts that is not stumped by a
situation wherein an entity that is neither a good instance of $A$ nor $B$
is nevertheless a good instance of $A$ AND $B$. Note that whereas in
representational approaches, relationships between concepts arise through
overlapping context-independent distributions, in the present approach, the
closeness of one concept to another (expressed as the probability that its
potentiality state will collapse to an actualized state of the other) is
context-dependent. Thus it is possible for two states to be far apart with
respect to a one context (for example $d_\mu(q, e, p)$, the distance
between {\bf guppy} and the cognitive state of the subject prior to the
context of being asked to name a pet), and close to one another with
respect to another context (for example $d_\mu(q, g, p)$, the distance
between {\bf guppy} and the cognitive state of the subject prior to the
context of being asked to name a pet fish).

\section{Summary and Conclusions}
Representational theories of concepts---such as prototype, exemplar, and
schemata or theory-based theories---have been adequate for describing
cognitive processes occurring in a focused, evaluative, analytic mode,
where one analyzes relationships of cause and effect. However, they have
proven to be severely limited when it comes to describing cognitive
processes that occur in a more intuitive, creative, associative mode, where
one is sensitive to and contextually responds to not just the most typical
properties of an item, but also less typical (and even hypothetical or
imagined) properties. This mode evokes relationships of not causation, but
correlation, such that new conjunctions of concepts emerge spontaneously.
This issue of conjunctions appears to have thrown a monkey wrench into
concepts research, but we see this as a mixed blessing. It brought to light
two things that have been lacking in this research: the notion of `state',
and a rigorous means of coping with potentiality and context.

First a few words about the notion of `state'. In representational
approaches, a concept is represented by one or more of its states. A
prototype, previously encountered exemplar, or theory description
constitutes merely one state of a concept. The competition between
different representational approaches seems to boil down to `which of the
states of a concept most fully captures the potentiality of the concept'?
Since different experimental designs elicit different context-specific
instantiations of a concept, it is not surprising that the states focused
on in one theory have greater predictive power in some experiments, while
the states focused on in another theory have greater predictive power in
others. The true state of affairs, however, is that none of the states can
represent the whole of the concept, just as none of the states of a
billiard ball can represent the whole of the billiard ball. The billiard
ball itself is described by the structure of the state space, which
includes all possible states, given the variables of interest and how they
could change. If one variable is location and another velocity, then each
location-velocity pair constitutes a state in this state space. To
represent the whole of an entity---whether it be a concept or a physical
object---one needs to consider the set of all states, and the structure
this set has.

This is the motivation for describing the essence of a concept as a
potentiality state. The potentiality state can, under the influence of a
context, collapse to a prototype, experienced exemplar, or an imagined or
counterfactual instance. The set of all these states, denoted $\Sigma_A$
for a concept $A \in {\cal A}$, is the state space of concept $A$. It is
this state space $\Sigma_A$, as a totality, together with the set of
possible contexts ${\cal M}$, and these two sets structured within the
$SCOP$ $(\Sigma_A, {\cal M}, \mu, {\cal L}, \nu)$ that represents the
concept. Hence a concept
is represented by an entire structure---including the possible states and
their properties, and the contexts that bring about change from one state
to another---rather than by one or a few specific state(s).

This brings us to the notion of context. If a theory is deficient with
respect to its consideration of state and state space, it is not unlikely
to be deficient with respect to the consideration of context, since
contexts require states upon which to act. The contextualized approach
introduced here makes use of a mathematical generalization of standard
quantum mechanics, the rationale being that the problems of context and
conjunction are very reminiscent to the problems of measurement and
entanglement that motivated the quantum formalism. Below we summarize how
these two problems manifest in the two domains of physics and cognition,
and how they are handled by quantum mechanics and its mathematical
generalizations.

\begin{itemize}
\item {\bf The Measurement Problem for Quantum Mechanics. }
\noindent
To know the state of a micro-entity, one must observe or measure some
property of it. However, the context of the measurement process
itself changes the state of the micro-entity from superposition state to an
eigenstate with respect to that measurement. Classical physics does not
incorporate a means of modeling change of state under the influence of
context. The best it can do is to avoid as much as possible any
influence of the measurement on the physical entity under study. However,
the change of
state under the influence of a measurement context---the quantum
collapse---is explicitly taken into account in the quantum mechanical
formalism. The state prior to, and independent of, the measurement, can be
retrieved as a theoretical object---the unit vector of complex
Hilbert space---that reacts to all possible measurement contexts in
correspondence with experimental results. Quantum
mechanics made it possible to describe the real
undisturbed and unaffected state of a physical entity even if most
of the experiments that are needed to measure properties of this entity
disturb this state profoundly (and often even destroy it).

\item {\bf The Measurement Problem for Concepts. }
\noindent
According to Rips' No Peeking Principle, we cannot be expected to
incorporate into a model of a concept how the concept interacts with
knowledge external to it. But {\it can} a concept be observed, studied, or
experienced in the absence of a context, something external to
it, whether that be a stimulus situation or another concept? We think not.
We adopt a Peeking Obligatory approach; concepts require a
peek---a measurement or context---to be elicited, actualized, or
consciously experienced. The generalization of quantum mechanics that we
use enables us to explicitly incorporate the context that elicits a
reminding of a concept, and the change of state this induces in the
concept, into the formal description of the concept itself. The concept in
its undisturbed state can then be `retrieved' as a superposition
of its instantiations. 

\item {\bf The Entanglement Problem for Quantum Mechanics. }
\noindent
Classical physics could successfully describe and predict relationships of
causation. However, it could not describe the correlations and
the birth of new states and new properties when micro-entities interact and
form a joint entity. Quantum mechanics describes this as a
state of entanglement, and use of the tensor product gives new states with
new properties.

\item {\bf The Entanglement Problem for Concepts. }
\noindent
Representational theories could successfully describe and predict the
results of cognitive processes involving relationships of
causation. However, they could not describe what happens when concepts
interact to form a conjunction, which often has properties that were
not present in its constituents. We treat conjunctions as concepts in the
context of one another, and we investigate whether the relative $SCOP$
might prove to be the algebraic
operation that corresponds to conjunction.
\end{itemize}
\noindent
Note that the measurement / peeking problem and the entanglement / conjunction
problem both involve context. The measurement / peeking problem concerns a
context
very external to, and of a
different sort from, the entity under consideration: an
observer or measuring apparatus in the case of physics, and a stimulus in
the case of cognition. In the entanglement / conjunction problem,
the context is the same sort of entity as the entity under consideration:
another particle in the case of physics, or another concept in
the case of cognition. The flip side of contextuality is potentiality; they
are two facets of the more general problem of describing the
kind of nondeterministic change of state that takes place when one has
incomplete knowledge of the universe in which the entity (or
entities) of interest, and the measurement apparatus, are operating.

The formalisms of quantum mechanics inspired the development of
mathematical generalizations of these formalisms such as the State COntext
Property system, or $SCOP$, with which one can describe situations of
varying degrees of
contextuality. In the $SCOP$ formalism, pure classical structure (no
effect of context) and pure quantum structure (completely contextual)
fall out as special cases. Applying the $SCOP$ formalism to concepts, pure
analytic (no effect of context) and pure associative (completely
contextual) modes fall out as special cases. In an analytic mode, cognitive
states consist of pre-established concepts. In an associative
mode, cognitive states are likely to be potentiality states ({\it i.e.} not
collapsed) with respect to contexts. This can engender a recursive
process in which the content of the cognitive state is repeatedly reflected
back at the associative network until it has been completely
defined in terms of some conjunction of concepts, and thus potentiality
gets reduced or eliminated with respect to the context. Eventually
a new stimulus context comes along for which this new state is a
superposition state, and the collapse process begins again. It has been
proposed that the onset of the capacity for a more associative mode of
thought is what lay behind the the origin of culture approximately
two million years ago (Gabora 1998, submitted), and that the capacity to
shift back and forth at will from analytical to associative
thought is what is responsible for the unprecedented burst of creativity in
the Middle/Upper Paleolithic (Gabora, submitted).

We suggest that the reason conjunctions of concepts can be treated as
entangled states is because of the presence of nonlocal EPR-type
correlations amongst the properties of concepts, which arise because they
exist in states of potentiality, with the presence or absence of
particular properties of a concept being determined {\it in the process of}
evoking or actualizing it. If, concepts are indeed entangled,
and thus for any concept, given the right context, any feature could
potentially become involved in its collapse, then the notion of
conceptual distance loses some meaning. What {\it can} be defined is not
the distance between concepts, but the distance between {\it
states} of them.\footnote{Note that this is also the case for physical
entities, even
in the macro-world described by classical physics.
One does not calculate the distance between two billiard balls, but rather
the distance between specific states of the two billiard
balls.} That said, the measure $d_\mu$ determines the distance between
the cognitive state prior to context (hence a potentiality state)
to the state after the influence of context (hence the collapsed state).
The measure $d_p$ determines distance between two potentiality
states. Note that the distance measures used in the prototype and exemplar
models are actually distances between states of concepts, not
between concepts themselves. This means that the distances we introduce are
no less fundamental or real as measures of conceptual
distance.

Preliminary theoretical evidence was obtained for the utility of the
approach, using the Pet Fish Problem. Conjunctions such as this are
dealt with by incorporating context-dependency, as follows: (1) activation
of {\bf pet} still rarely causes activation of {\bf guppy}, and
likewise (2) activation of {\bf fish} still rarely causes activation of
{\bf guppy}. But now (3) {\bf pet fish} causes activation of the
superposition state {\bf pet} {\it in the context of} {\bf pet fish} AND
{\bf fish} {\it in the context of} {\bf pet fish}. Since for this
superposition state the probability of collapsing onto the state {\bf
guppy} is high, it is very likely to be activated. Thus we have a
formalism for describing concepts that is not stumped by the sort of
widespread anomalies that arise with concepts, such as this situation
wherein an entity that is neither a good instance of $A$ nor $B$ is
nevertheless a good instance of the conjunction of $A$ and $B$.

Despite our critique of representational approaches, the approach
introduced here was obviously derived from and inspired by them. Like
exemplar theory, it emphasizes the capacity of concepts to be instantiated
as different exemplars. In agreement to some extent with
prototype theory, experienced exemplars are `woven together', though
whereas a prototype is limited to some subset of all conceivable
features, a potentiality state is not. Our way of dealing with the
`insides' of a concept is more like that of the theory or schemata
approach. An instance is described as, not a set of weighted features, but
a lattice that represents its relational structure. The
introduction of the notion of a concept core, and the return of the notion
of essence, have been useful for understanding how what is most
central to a concept could remain unscathed in the face of modification to
the concept's mini-theory. Our distinction between state of
instantiation and potentiality state is reminiscent of the distinction
between theory and core. However, the introduction of a core cannot
completely rescue the theory theory until serious consideration has been
given to state and context.

We end by asking: does the contextualized approach introduced here bring us
closer to an answer to the basic question `what is a concept'?
We have sketched out a theory in which concepts are not fixed
representations but entities existing in states of potentiality that get
dynamically actualized, often in conjunction with other concepts, through a
collapse event that results from the interaction between
cognitive state and stimulus situation or context. But does this tell us
what a concept really is? Just as was the case in physics a
century ago, the quantum formalism, while clearing out many troubling
issues, confronts us with the limitations of science. We cannot step
outside of any particular orientation and observe directly and objectively
what a concept is. The best we can do is reconstruct a concept's
essence from the contextually elicited `footprints' it casts in the
cognitive states that make up a stream of thought.

\section*{Appendices}

\appendix

\section{Complex Hilbert Space} \label{appendix01}
A complex Hilbert space ${\cal H}$ is a set such that for two elements $x,
y \in {\cal H}$ of this set an operation
`sum' is defined, denoted $x+y$, and for any element $x \in {\cal H}$ and
any complex number $\lambda \in \compl$, the
multiplication of this element $x$ with this complex number $\lambda$ is
defined, denoted by $\lambda x$. The operation
`sum' and `multiplication by a complex number' satisfy the normal
properties that one expect these operations to
satisfy ({\it e.g.} $x+y=y+x$, $(x+y)+z=x+(y+z)$, $\lambda \mu x = \mu
\lambda x$, {\it etc}, \ldots
A complete list of all these properties can be found in any textbook on
vector spaces). So this makes the
set ${\cal H}$ into a complex vector space, and thus we call
the elements $x \in {\cal H}$ vectors.

In additional to the two operations of `sum' and `multiplication by a
complex number', a Hilbert space has an operation
that is called the `inproduct of vectors'. For two vectors $x, y \in {\cal
H}$ the inproduct is denoted $\langle x, y
\rangle$, and it is a complex number that has the following properties. For
$x, y, z \in {\cal H}$, and $\lambda \in
\compl$, we have
\bea
\langle x, y \rangle &=& \langle y, x \rangle^* \\
\langle x , y + \lambda z \rangle &=& \langle x, y \rangle+\lambda \langle
x, z \rangle
\eea
The inproduct makes it possible to define an orthogonality relation on the
set of vectors. Two vectors $x, y \in {\cal
H}$ are orthogonal, and we denote $x \perp y$, if and only if $\langle x, y
\rangle = 0$. Suppose that we consider a
subset $A \subset {\cal H}$, then we can introduce
\be
A^\perp = \{x\ \vert\ x \in {\cal H}, x \perp y\ \forall\ y \in A\}
\ee
which consists of all the vectors orthogonal to all vectors in $A$. It is
easy to verify that $A^\perp$ is a
subspace of ${\cal H}$, and we call it the orthogonal subspace to $A$. We
can also show that $A \subset (A^\perp)^\perp$, and
call $(A^\perp)^\perp$, also denoted $A{^\perp}{^\perp}$, the biorthogonal
subspace of $A$.

There is one more property satisfied to make the complex vectorspace with
an inproduct into a Hilbert space, and that
is, for $A \subset {\cal H}$ we have:
\be
A^\perp + A{^\perp}{^\perp} = {\cal H}
\ee
This means that for any subset $A \subset {\cal H}$, each vector $x \in
{\cal H}$ can always be written as the
superposition
\be
x = y + z \label{eq:superposition}
\ee
where $y \in A^\perp$ and $z \in A{^\perp}{^\perp}$.
The inproduct also introduces for two vectors $x, y \in {\cal H}$ the
measure of a distance and an angle between these two
vectors as follows:
\bea
d(x, y) &=& \sqrt{\langle x-y, x-y \rangle} \label{eq:Hilbertdistance} \\
\theta &=& \arccos|\langle x, y \rangle| \label{eq:Hilbertangle}
\eea
and for one vector $x \in {\cal H}$, the measure of a length of this vector
\be
\|x\| = \sqrt{\langle x, x \rangle}
\ee
This distance makes the Hilbert space a
topological space (a metric space). It can be shown that for
$A
\subset {\cal H}$ we have that $A^\perp$ is a topologically closed subspace
of ${\cal H}$, and that the biorthogonal operation
is a closure operation. Hence
$A{^\perp}{^\perp}$ is the closure of
$A$. This completes the mathematical definition of a complex Hilbert space.

\section{Quantum Mechanics in Hilbert space} \label{appendix02}
In quantum mechanics, the states of the physical entity under study are
represented by the unit vectors of a complex Hilbert
space ${\cal H}$. Properties are represented by closed subspaces of ${\cal
H}$, hence subsets that are of the form
$A{^\perp}{^\perp}$ for some $A \subset {\cal H}$. Let us denote such
closed subspaces by $M \subset {\cal H}$, and the
collection of all closed subspaces by ${\cal P}({\cal H})$. For a physical
entity in a state $x \in {\cal H}$, where $x$ is a
unit vector, we have that property $M$ is `actual' if and inly if $x \in
M$. Suppose that we consider a physical entity in a
state $x \in {\cal H}$ and a property $M \in {\cal P}({\cal H})$ that is
not actual, hence potential. Then, using
(\ref{eq:superposition}), we can determine the weight of this property.
Indeed there exists vectors $y, z \in {\cal H}$ such
that
\be
x = y + z
\ee
and $y \in M$ and $z \in M^\perp$. We call the vector $y$ the projection of
$x$ on $M$, and denote it $P_M(x)$, and the vector
$z$ the projection of $x$ on $M^\perp$, and denote it $P_{M^\perp}(x)$. The
weight $\nu(x, M)$ of the property $M$ for the state
$x$ is then given by
\be
\nu(x, M) = \langle x, P_M(x) \rangle \label{eq:weight}
\ee
The vectors $y / \|y\|$ (or $P_M(x) / \|P_M(x)\|$) and $z / \|z\|$ (or
$P_{M^\perp}(x) / \|P_{M^\perp}(x)\|$) are also called the collapsed
vectors under measurement context $\{M, M^\perp\}$. An
arbitrary measurement context $e$ in quantum mechanics is represented by a
set of closed subspaces $\{M_1, M_2, \ldots, M_n,
\ldots\}$ (eventually infinite), such that
\bea
&M_i \perp M_j\ \forall\ i \not= j \\
&\sum_iM_i = {\cal H}
\eea
The effect of such a measurement context $\{M_1, M_2, \ldots, M_n,
\ldots\}$ is that the state $x$ where the physical entity is
in when the measurement context is applied collapses to one of the states
\be
P_{M_i}(x) \over \|P_{M_i}(x)\|
\ee
and the probability $\mu(P_{M_i}(x), e, x)$ of this collapse is
given by
\be
\mu(P_{M_i}(x), e, x) = \langle x, P_{M_i}(x) \rangle
\label{eq:collapseprobability}
\ee
If we compare (\ref{eq:weight}) and (\ref{eq:collapseprobability}) we see
that for a quantum mechanical entity the weight of a
property $M$ for a state $x$ is equal to the probability that the state $x$
will collapse to the state $P_M(x) / \|P_M(x)\|$,
if the measurement context $\{M, M^\perp\}$ is applied to this physical
entity in this state. That is the reason that it would be interesting to
compare these quantities in the
case of concepts (see section
\ref{sec:weightprobability}).

\section{SCOP Systems of Pure Quantum Mechanics}
\label{appendix03}
The set of states $\Sigma_Q$ of a quantum entity is the set of unit vectors
of the complex Hilbert space ${\cal H}$. The set of
contexts ${\cal M}_Q$ of a quantum entity is the set of measurement
contexts, {\it i.e.} the set of sequences $\{M_1, M_2,
\ldots, M_n, \dots\}$ of closed subspaces of the Hilbert space ${\cal H}$,
such that
\bea
&M_i \perp M_j\ \forall\ i \not= j \\
&\sum_iM_i = {\cal H}
\eea
Such sequence is also called a {\it spectral family}. The word spectrum
refers to the set of possible
outcomes of the measurement context under consideration. In quantum
mechanics, a state $p \in \Sigma_Q$ changes to another
state $q \in \Sigma_Q$ under influence of a context $e \in {\cal M}_Q$ in
the following way. If $\{M_1, M_2,
\ldots, M_n, \dots\}$ is the spectral family representing the context $e$,
and $x$ the unit vector representing the state $p$,
then $q$ is one of the unit vectors
\be
P_{M_i}(x) \over \|P_{M_i}(x)\|
\ee
and the change of $x$ to $P_{M_i}(x) / \|P_{M_i}(x)\|$ is called the
quantum collapse. The probability of this change is given by
\be
\mu(e, q, e, p) = \langle x, P_{M_i}(x) \rangle
\ee
Remark that in quantum mechanics the context $e$ is never changed. This
means that
\be
\mu(f, q, e, p) = 0\ \forall f \not= e
\ee
As a consequence, we have for the reduced probability (see
(\ref{eq:reducedprobability}))
\be
\mu(q, e, p) = \mu(e, q, e, p) = \langle x, P_{M_i}(x) \rangle
\ee
A property $a$ of a quantum entity is represented by a closed subspace $M$
of the complex Hilbert space ${\cal H}$. A property
$a$ represented by $M$ always has a unique orthogonal property $a^\perp$
represented by $M^\perp$ the orthogonal closed subspace
of
$M$. This orthogonal property $a^\perp$ is the quantum-negation of the
property $a$. The weight $\nu(p, a)$ of a property $a$
towards a state
$p$ is given by
\be
\nu(p, a) = \langle x, P_{M_i}(x) \rangle
\ee
where $M$ represents $a$ and $x$ represents $p$. Remark that at first sight,
the weight does not appear to depend on a context, as it does for a
general state context property system. This is only partly
true. In pure quantum mechanics, the weights only
depend on context in an indirect way, namely because a property introduces
a unique context, the context corresponding to the
measurement of this property. This context is represented by the spectral
family $\{M, M^\perp\}$.

\section{SCOP Systems Applied to Cognition}
\label{appendix04}
A state context property system $(\Sigma, {\cal M}, {\cal L}, \mu, \nu)$
consists of three sets $\Sigma, {\cal M}$ and ${\cal L}$,
and two functions $\mu$ and $\nu$.

$\Sigma$ is the set of cognitive states of the subjects under
investigation, while ${\cal M}$ is
the set of contexts that influence and change these cognitive states.
${\cal L}$ represents properties or features of concepts. The
function $\mu$ is defined from the set ${\cal M} \times \Sigma \times {\cal
M} \times \Sigma$ to the interval $[0, 1]$ of real
numbers, such that
\be
\sum_{f \in {\cal M}, q \in \Sigma}\mu(f, q, e, p) = 1
\ee
and $\mu(f, q, e, p)$ is the probability that the cognitive state $p$
changes to cognitive state $q$ under influence of context $e$
entailing a new context $f$.

We noted that properties of concepts can also be treated as concepts.
Remark also that it often makes sense to treat concepts as features. For
example, if we say `a dog is an animal', it is in fact the feature `dog' of
the object in front of us that we relate to the feature `animal' of this
same physical object. This means that a relation like `dog is animal' can
be expressed within the structure ${\cal L}$ in our formalism.

This relation is the first structural element of the set ${\cal L}$, namely
a partial order relation, denoted $<$. A property $a \in {\cal L}$
`implies' a property $b \in {\cal L}$, and we denote $a < b$, if and only
if, whenever $a$ is true then also $b$ is true. This partial order relation
has the following properties. For $a, b \in {\cal L}$ we have:
\bea
&a < a \\
&a < b\ {\rm and}\ b < a \Rightarrow a = b \\
&a < b\ {\rm and}\ b < c \Rightarrow a < c
\eea
For a set of properties $\{a_i\}$ there exists a conjunction property
denoted $\wedge_ia_i$. This conjunction property $\wedge_ia_i$ is true if
and only if all of the properties $a_i$ are true. This means that for $a_i,
b \in {\cal L}$ we have:
\be
b < \wedge_ia_i \Leftrightarrow b < a_i\ \forall i
\ee
The conjunction property defines mathematically an infimum for the partial
order relation $<$. Hence we demand that each subset of ${\cal L}$ has an
infimum in ${\cal L}$, which makes ${\cal L}$ into a {\it complete lattice}.

Each property $a$ also has the `not' (negation) of this property, which we
denote $a^\perp$. This is mathematically expressed by demanding
that the lattice ${\cal L}$ be equipped with an orthocomplementation, which
is a function from ${\cal L}$ to ${\cal L}$ such that for $a, b
\in {\cal L}$ we have:
\bea
&(a^\perp)^\perp = a \\
&a < b \Rightarrow b^\perp < a^\perp \\
&a \wedge a^\perp = 0
\eea
where $0$ is the minimal property (the infimum of all the elements of
${\cal L}$), hence a property that is never true. This makes ${\cal
L}$ into a complete orthocomplemented lattice.

The function $\nu$ is defined from the set $\Sigma \times {\cal M} \times
{\cal L}$ to the interval
$[0, 1]$, and $\nu(p, e, a)$ is the weight of property $a$ under context
$e$ for state $p$. For $a \in {\cal L}$ we have:
\be
\nu(p, e, a) + \nu(p, e, a^\perp) = 1
\ee

\section*{Acknowledgements}

We would like to acknowledge the
support of Grant G.0339.02 of the Flemish Fund for Scientific Research.

\section*{References}

Abeles, M. \& Bergman, H., 1993, Spatiotemporal firing patterns in the frontal cortex of behaving monkeys.
{\it Journal of Neurophysiology} {\bf 70}(4): 1629-1638.
\medskip
\noindent
Aerts, D., 1982, Description of many physical entities without the
paradoxes encountered in quantum mechanics.
{\it Foundations of Physics} {\bf 12}: 1131-1170.

\medskip
\noindent
Aerts, D., 1983, Classical theories and nonclassical theories as a
special case of a more
general theory. {\it Journal of Mathematical Physics} {\bf 24}: 2441-2453.

\medskip
\noindent
Aerts, D., 1985, The physical origin of the EPR paradox and how to violate
Bell inequalities by
macroscopical systems. In P. Lathi and P. Mittelstaedt (eds) {\it On the
Foundations of Modern Physics} (World Scientific: Singapore), pp. 305-320.

\medskip
\noindent
Aerts, D., 1991, A mechanistic classical laboratory situation violating
the Bell inequalities
with $\sqrt{2}$, exactly `in the same way' as its violations by the EPR
experiments.
{\it Helvetica Phyica Acta} {\bf 64}: 1-23.

\medskip
\noindent
Aerts, D., 1993, Quantum structures due to fluctuations of the
measurement situations.
{\it International Journal of Theoretical Physics} {\bf 32}: 2207-2220.

\medskip
\noindent
Aerts, D., 2002, Being and change: foundations of a realistic operational
formalism. In D. Aerts {\it et al.} (eds) {\it Probing the Structure of
Quantum Mechanics: Nonlinearity, Nonlocality,
Computation and Axiomatics} (World Scientific, Singapore).

\medskip
\noindent
Aerts, D. and Aerts, S., 1994, Applications of quantum statistics in psychological studies of decision processes, {\it Foundations of Science} {\bf 1}: 85-97.

\medskip
\noindent
Aerts, D., Aerts, S., Broekaert, J. and Gabora, L., 2000a, The violation
of Bell inequalities in
the macroworld. {\it Foundations of Physics} {\bf 30}: 1387-1414.

\medskip
\noindent
Aerts, D., Broekaert, J. and Gabora, L., 1999, Nonclassical contextuality
in cognition:
borrowing from quantum mechanical approaches to indeterminism and observer
dependence. In R. Campbell (ed.) {\it Dialogues Proceedings of Mind IV
Conference}, Dublin, Ireland.

\medskip
\noindent
Aerts, D., Broekaert, J. and Gabora, L., 2000b, Intrinsic contextuality as
the crux of
consciousness. In K. Yasue (ed.) {\it Fundamental Approaches to
Consciousness}. (Amsterdam: John Benjamins
Publishing Company), pp. 173-181.

\medskip
\noindent
Aerts, D. and Durt, T., 1994a, Quantum, classical and intermediate: a
measurement model. In C. Montonen (ed.)
{\it Proceedings of the International Symposium on the Foundations of
Modern Physics 1994},
Helsinki, Finland, (Editions Frontieres: Gives Sur
Yvettes, France).

\medskip
\noindent
Aerts, D., Colebunders, E., Van der Voorde, A. and Van Steirteghem, B.,
1999, State property
systems and closure spaces: A study of categorical equivalence. {\it
International Journal of
Theoretical Physics} {\bf 38}: 359-385.

\medskip
\noindent
Aerts, D. and Durt, T., 1994b, Quantum, classical, and intermediate, an
illustrative example.
{\it Foundations of Physics} {\bf 24}: 1353-1368.

\medskip
\noindent
Aerts, D., D'Hondt, E. and Gabora, L., 2000c, Why the disjunction in
quantum logic is not
classical. {\it Foundations of Physics} {\bf 30}: 1473-1480.

\medskip
\noindent
Aerts, D. and Gabora, L., 2002, Toward a state and context theory of
concepts. In preparation.

\medskip
\noindent
Bell, J., 1964, On the Einstein Podolsky Rosen paradox. {\it Physics}
{\bf 1}: 195.

\medskip
\noindent
Boden, M., 1991, {\it The Creative Mind: Myths and Mechanisms}. (Cambridge
UK: Cambridge University Press).

\medskip
\noindent
Campbell, D., 1987, Evolutionary epistemology. In G. Radnitzky and W. W.
Bartley III, (eds) {\it Evolutionary Epistemology, Rationality, and the Sociology of Knowledge} (La Salle,
Illinois: Open Court).

\medskip
\noindent
Campbell, F. W. \& Robson, J. G., 1968, Application of Fourier analysis to the visibility of gratings. {\it
Journal of Physiology} {\bf 197}: 551-566.

\medskip
\noindent
Cariani, P., 1995, As if time really mattered: temporal strategies for neural coding of sensory information. In {\it Origins: Brain and self-organization}, ed. K. Pribram. (Hillsdale NJ: Erlbaum), pp. 161-229.

\medskip
\noindent
Cariani, P., 1997, Temporal coding of sensory information. In {\it Computational neuroscience: Trends in research 1997}, ed. J.M. Bower. (Dordrecht, Netherlands: Plenum), pp. 591-598.

\medskip
\noindent
Dartnell, T. 1993, Artificial intelligence and creativity: an
introduction. {\it Artificial
Intelligence and the Simulation of Intelligence Quarterly} {\bf 85}.

\medskip
\noindent
Dennett, D., 1987, {\it Brainstorms: Philosophical Essays on Mind and
Psychology} (Harvester Press).

\medskip
\noindent
Dewing, K. \& Battye, G., 1971, Attentional deployment and non-verbal fluency. {\it Journal of Personality and
Social Psychology}, 17: 214-218.

\medskip
\noindent
De Valois, R.L. \& De Valois, K. K., 1988, {\it Spatial Vision} (Oxford UK: Oxford University Press).

\medskip
\noindent
Dykes, M. \& McGhie, M., 1976, A. A comparative study of attentional strategies in schizophrenics and highly
creative normal subjects. {\it British Journal of Psychiatry} 128: 50-56.

\medskip
\noindent
Emmers, R., 1981, {\it Pain: A Spike-Interval Coded Message in the Brain} (Philadelphia PA: Raven Press).

\medskip
\noindent
Fodor, J. A. 1994, Concepts: A pot-boiler. {\it Cognition} {\bf 50}: 95-113.

\medskip
\noindent
Foulis, D., Piron C., and Randall, C., 1983, Realism, operationalism and
quantum mechanics.
{\it Foundations of Physics} {\bf 13}.

\medskip
\noindent
Foulis, D. and Randall, C., 1981, What are quantum logics and what ought
they to be? In E. Beltrametti, and B. van Fraassen (eds) {\it Current
Issues in Quantum Logic}, {\bf 35} (New York: Plenum Press).

\medskip
\noindent
Gabora, L., 1998, Autocatalytic closure in a cognitive system: A tentative
scenario for the origin of culture. {\it Psycholoquy} {\bf 9}(67).
\\ http://www.cogsci.soton.ac.uk/cgi/psyc/newpsy?9.67 [adap-org/9901002]

\medskip
\noindent
Gabora, L., 2000, Toward a theory of creative inklings. In R. Ascott (ed)
{\it Art, Technology, and
Consciousness}, (Bristol UK: Intellect Press), pp. 159-164.
http://cogprints.soton.ac.uk/documents/disk0/00/00/08/56/

\medskip
\noindent
Gabora, L., 2002a, The beer can theory of creativity. In P. Bentley and D.
Corne (eds.) {\it Creative
Evolutionary Systems}, (San Francisco: Morgan Kauffman), pp. 147-161.
\\ http://cogprints.soton.ac.uk/documents/disk0/00/00/09/76/

\medskip
\noindent
Gabora, L., 2002b, Cognitive mechanisms underlying the creative process.
{\it Proceedings of the Fourth International Workshop on Creativity and
Cognition}.
http://www.vub.ac.be/CLEA/liane/papers/cogmech/cogmech.htm

\medskip
\noindent
Gabora, L., Origin of the modern mind through conceptual closure, submitted.
http://www.vub.ac.be/CLEA/liane/papers/ommcc/ommcc.htm

\medskip
\noindent
Gerrig, R. and Murphy, G., 1992, Contextual Influences on the
Comprehension of Complex concepts.
{\it Language and Cognitive Processes} {\bf 7}: 205-230.

\medskip
\noindent
Goldstone, R. L., and Rogosky, B. J. (in press). The role of roles in
translating across conceptual systems. {\it Proceedings of the
Twenty-fourth Annual Conference of the Cognitive Science Society.}
(Hillsdale, New Jersey: Lawrence Erlbaum Associates).

\medskip
\noindent
Hampton, J., 1987, Inheritance of attributes in natural concept
conjunctions. {\it Memory \&
Cognition} {\bf 15}: 55-71.

\medskip
\noindent
Hampton, J., 1997, Conceptual combination. In Lamberts and D. Shanks (eds)
{\it Knowledge, Concepts, and Categories} (Psychology Press).

\medskip
\noindent
Hancock, P. J. B., Smith, L. S. and Phillips, W. A., 1991, A biologically
supported error-correcting
learning rule. {\it Neural Computation} {\bf 3}(2): 201-212.

\medskip
\noindent
Hastie, R., Schroeder, C., and Weber, R., 1990, Creating complex social
conjunction categories from simple
categories. {\it Bulletin of the Psychonomic Society}, {\bf 28}: 242-247.

\medskip
\noindent
Hebb, D. O., 1949, {\it The Organization of Behavior} (Wiley).

\medskip
\noindent
Heit, E. and Barsalou, L., 1996, The instantiation principle in natural
language categories.
{\it Memory} {\bf 4}: 413-451.

\medskip
\noindent
Holden, S.B. and Niranjan, M., 1997, Average-case learning curves for
radial basis function
networks. {\it Neural Computation} {\bf 9}(2): 441-460.

\medskip
\noindent
James, W., 1890/1950, {\it The Principles of Psychology} (New York: Dover).

\medskip
\noindent
Jauch, J., 1968, {\it Foundations of Quantum Mechanics} (Reading Mass:
Addison-Wesley).

\medskip
\noindent
Johnson-Laird, P. N., 1983, {\it Mental Models}. (Cambridge Mass: Harvard
University Press).

\medskip
\noindent
Komatsu, L., 1992, Recent views of conceptual structure. {\it Psychological
Bulletin} {\bf 112}: 500-526.

\medskip
\noindent
Kunda, Z., Miller, D.T., and Clare, T., 1990, Combining social concepts:
the role of causal reasoning. {\it Cognitive Science} {\bf 14}: 551-578.

\medskip
\noindent 
Lestienne, R., 1996, Determination of the precision of spike timing in the visual cortex of anesthetized cats. Biology and Cybernetics, 74: 55-61.
\medskip
\noindent 
Lestienne, R. amd Strehler, B. L., 1987, Time structure and stimulus dependence of precise replicating patterns present in monkey cortical neuron spike trains. Neuroscience (April).

\medskip
\noindent
Lu, Y.W., Sundararajan, N. and Saratchandran, P. 1997, A sequential learning scheme for function approximation using minimal radial basis function neural networks. {\it Neural Computation} {\bf 9}(2): 461-478.

\medskip
\noindent
Mackey, G., 1963. {\it Mathematical Foundations of Quantum Mechanics}
(Reading Mass: Benjamin).

\medskip
\noindent 
Martindale, C., 1977, Creativity, consciousness, and cortical arousal. Journal of Altered States of Consciousness, 3: 69-87.

\medskip
\noindent
Marr, D., 1969. A theory of the cerebellar cortex. {\it Journal of
Physiology} {\bf 202}: 437-470.

\medskip
\noindent
Martindale, C., 1999, Biological bases of creativity. In Handbook of Creativity, ed. R. J. Sternberg, Cambridge University Press, Cambridge UK, 137-152.

\medskip
\noindent
Martindale, C. \& Armstrong, J., 1974, The relationship of creativity to cortical activation and its operant
control. Journal of Genetic Psychology, 124: 311-320.

\medskip
\noindent
Medin, D., Altom, M., and Murphy, T., 1984, Given versus induced category
representations: Use of
prototype and exemplar information in classification. {\it Journal of
Experimental Psychology:
Learning, Memory, and Cognition} {\bf 10}: 333-352.

\medskip
\noindent
Medin, D. and Ortony, A., 1989, Psychological essentialism. In S. Vosniadou
and A. Ortony (eds) {\it Similarity and Analogical Reasoning.} (Cambridge:
Cambridge University Press) pp. 179-195.

\medskip
\noindent
Medin, D. and Schaffer, M. M., 1878, A context theory of classification
learning. {\it Psychological Review} {\bf 85}: 207-238.

\medskip
\noindent
Medin, D. and Shoben, E., 1988, Context and Structure in Conceptual
Combinations. {\it Cognitive Psychology} {\bf 20}: 158-190.

\medskip
\noindent
Mendelsohn, G. A., 1976, Associative and attentional processes in creative performance. {\it Journal of Personality} {\bf 44}: 341-369.

\medskip
\noindent
Murphy, G. and Medin, D., 1985, The role of theories in conceptual
coherence. {\it Psychological Review} {\bf 92}: 289-316.

\medskip
\noindent
Neisser, U., 1963, The multiplicity of thought. {\it British Journal of
Psychology} {\bf 54}: 1-14.

\medskip
\noindent
Nosofsky, R., 1988, Attention, similarity, and the
identification-categorization relationship. {\it Journal of Experimental
Psychology:
General} {\bf 115}: 39-57.

\medskip
\noindent
Nosofsky, R., 1988, Exemplar-based accounts of relations between
classification, recognition, and typicality. {\it Journal of Experimental
Psychology: Learning, Memory, and Cognition} {\bf 14}: 700-708.

\medskip
\noindent
Nosofsky, R., 1992, Exemplars, prototypes, and similarity rules. In A.
Healy, S. Kosslyn, and
R. Shiffrin (eds) {\it From Learning Theory to
Connectionist Theory: Essays in Honor of William K. Estes} {\bf 1}
(Hillsdale NJ: Lawrence Erlbaum), pp. 149-167.

\medskip
\noindent
Osherson, D. and Smith, E., 1981, On the adequacy of prototype theory as a
theory of concepts.
{\it Cognition} {\bf 9}: 35-58.

\medskip
\noindent
Piaget, J., 1926, {\it The Language and Thought of the Child} (London:
Routledge \& Kegan Paul).

\medskip
\noindent
Piron, C., 1976, {\it Foundations of Quantum Physics} (Reading Mass: W. A.
Benjamin).

\medskip
\noindent
Piron, C., 1989, Recent developments in quantum mechanics. {\it Helvetica
Physica Acta},
{\bf 62}: 82.

\medskip
\noindent
Piron, C., 1990, {\it M\'ecanique Quantique: Bases et Applications},
(Lausanne: Press Polytechnique de
Lausanne).

\medskip
\noindent
Pitowsky, I., 1989, {\it Quantum Probability - Quantum Logic}, Lecture Notes
in Physics 321 (Berlin:
Springer).

\medskip
\noindent
Randall, C. and Foulis, D., 1976, A mathematical setting for inductive
Reasoning. In C. Hooker (ed.) {\it Foundations of Probability Theory,
Statistical Inference, and
Statistical Theories of Science
III}, 169, (Dordrecht: Kluwer Academic).

\medskip
\noindent
Randall, C. and Foulis, D., 1978, The Operational Approach to Quantum
Mechanics. In C. Hooker (ed.) {\it Physical
Theories as Logico-Operational Structures}, 167 (Dordrecht,
Kluwer Academic).

\medskip
\noindent
Reed, S., 1972, Pattern recognition and categorization. {\it Cognitive
Psychology} {\bf 3}:382-407.

\medskip
\noindent
Riegler, A. Peschl, M. and von Stein, A., 1999, {\it Understanding
Representation in the Cognitive
Sciences} (Dordrecht Holland: Kluwer Academic).

\medskip
\noindent
Rips, L.J., 1995, The current status of research on concept combination.
{\it Mind and Language}, {\bf 10}: 72-104.

\medskip
\noindent
Rips, L. J., 2001, Two kinds of reasoning. {\it Psychological Science} {\bf
12}: 129-134.

\medskip
\noindent
Rips, L. J., 2001, Necessity and natural categories. {\it Psychological
Bulletin} {\bf 127}(6): 827-852.

\medskip
\noindent
Rosch, E., 1975, Cognitive reference points. {\it Cognitive Psychology}
{\bf 7}: 532-547.

\medskip
\noindent
Rosch, E., 1978, Principles of categorization. In E. Rosch and
B. Lloyd (eds) {\it Cognition and
Categorization} (Hillsdale, NJ: Erlbaum). pp. 27-48.

\medskip
\noindent
Rosch, E., 1983, Prototype classification and logical classification: The
two systems. In E. K. Scholnick (ed.) {\it New Trends in Conceptual
Representation: Challenges to Piaget's Theory?} (Hillsdale, NJ: Erlbaum),
pp. 73-86.

\medskip
\noindent
Rosch, E., and Mervis, C., 1975, Family resemblances: Studies in the
internal structure of categories. {\it Cognitive Psychology} {\bf 7}: 573-605.

\medskip
\noindent
Rosch, E., 1999, Reclaiming concepts. {\it Journal of Consciousness
Studies} {\bf 6}(11): 61-78.

\medskip
\noindent
Rumelhart, D. E. and Norman, D. A., 1988, Representation in memory. In R.
C. Atkinson, R. J. Hernsein, G. Lindzey, and R. D. Luce (eds) Stevens'
Handbook of Experimental Psychology (New York: Wiley).

\medskip
\noindent
Sloman, S., 1996, The empirical case for two systems of Reasoning. {\it
Psychological Bulletin}
{\bf 9}(1): 3-22.

\medskip
\noindent
Smith, E., and Medin, D., 1981, {\it Categories and Concepts}. Cambridge MA:
Harvard University Press.

\medskip
\noindent
Storms, G., De Boeck, P., Hampton, J., and Van Mechelen, I., 1999,
Predicting conjunction
typicalities by component typicalities. {\it Psychonomic Bulletin and
Review} {\bf 4}: 677-684.

\medskip
\noindent
Storms, G., De Boeck, P., and Ruts, W., 2000, Prototype and exemplar based
information in natural language categories. {\it Journal of Memory and
Language} {\bf 42}: 51-73.

\medskip
\noindent
Storms, G., De Boeck, P., Van Mechelen, I. and Ruts, W., 1996, The
dominance effect in concept conjunctions: Generality and interaction
aspects. {\it Journal of Experimental Psychology: Learning,
Memory \& Cognition} {\bf 22}: 1-15.

\medskip
\noindent
Storms, G., De Boeck, P., Van Mechelen, I. and Ruts, W., 1998, Not
Guppies, nor goldfish, but tumble dryers, Noriega, Jesse Jackson, panties,
car crashes, bird books, and Stevie Wonder. {\it Memory \& Cognition} {\bf
26}: 143-145.

\medskip
\noindent
Sutcliffe, J., 1993, Concepts, class, and category in the tradition of
Aristotle. In I. Van Mechelen, J. Hampton, R. Michalski, and P. Theuns
(eds.) {\it Categories and Concepts: Theoretical Views and Inductive Data
Analysis}, (London: Academic Press), pp. 35-65.

\medskip
\noindent
Verbeemen, T., Vanoverberghe, V., Storms, G., and Ruts, W., 2002, The role
of contrast categories in
natural language concepts. Forthcoming.

\medskip
\noindent
Willshaw, D. J. and Dayan, P., 1990, Optimal plasticity from matrix
memory: What goes up must come down. {\it Journal of Neural Computation}
{\bf 2}: 85-93.

\medskip
\noindent
Wisniewski, E., 1997, When concepts combine. {\it Psychonomic Bulletin \&
Review} {\bf 4}: 167-183.

\medskip
\noindent
Wisniewski, E. and Gentner, D., 1991, On the combinatorial semantics of
noun pairs: Minor and major adjustments. In G. B. Simpson (ed.) {\it
Understanding Word and Sentence} (Amsterdam: Elsevier) pp. 241-284.

\end{document}